\documentclass{emulateapj}
\pdfoutput=1
\usepackage{apjfonts}
\usepackage{amsmath}
\usepackage{multirow}
\newcommand{\beq}{\begin{equation}}
\newcommand{\eeq}{\end{equation}}
\newcommand{\bea}{\begin{eqnarray}}
\newcommand{\eea}{\end{eqnarray}}

\begin{document}

\title{Nuclear Star Clusters from Clustered Star Formation}
\author{
Meghann Agarwal \altaffilmark{1} 
and
Milo\v s Milosavljevi\'c \altaffilmark{2}
}
\affil{\altaffilmark{1} Department of Physics, University of Texas, 1
  University Station C1600, Austin, TX 78712.
\\
\altaffilmark{2} Department of Astronomy and Texas Cosmology Center,
University of Texas, 1 University Station C1400, Austin, TX 78712.}
\righthead{NUCLEAR CLUSTERS FROM CLUSTERED STAR FORMATION}
\lefthead{AGARWAL \& MILOSAVLJEVI\'C}

\begin{abstract}

Photometrically distinct nuclear star clusters (NSCs) are common in
late-type-disk and spheroidal galaxies. The formation of NSCs is
inevitable in the context of normal star formation in which a majority
of stars form in clusters. A young, mass-losing cluster embedded in an
isolated star-forming galaxy remains gravitationally bound over a
period determined by its initial mass and the galactic tidal
field. The cluster migrates radially toward the center of the galaxy
and becomes integrated in the NSC if it reaches the center. The rate
at which the NSC grows by accreting young clusters can be estimated
from empirical cluster formation rates and dissolution times. We model
cluster migration and dissolution and find that the NSCs in late-type
disks and in spheroidals could have assembled from migrating
clusters. The resulting stellar nucleus contains a small fraction of
the stellar mass of the galaxy; this fraction is sensitive to the
high-mass truncation of the initial cluster mass function (ICMF).  The
resulting NSC masses are consistent with the observed values, but
generically, the final NSCs are surrounded by a spatially more
extended excess over the inward-extrapolated exponential (or S\'ersic)
law of the outer galaxy. We suggest that the excess can be related to
the pseudobulge phenomenon in disks, though perhaps not all of the
pseudobulge mass assembles this way. Comparison with observed NSC
masses can be used to constrain the truncation mass scale of the ICMF
and the fraction of clusters suffering prompt dissolution.  We infer
truncation mass scales of $\lesssim 10^6\,M_\odot$ ($\gtrsim
10^5\,M_\odot$) without (with $90\%$) prompt dissolution.  Since the
NSC assembly is collisionless and non-dissipative, no relation to the
process responsible for central black hole assembly in more massive
galaxies is expected.


\end{abstract}

\section{Introduction}
\setcounter{footnote}{0}

Photometrically distinct nuclear star clusters (NSCs) are ubiquitous
in dynamically primitive galaxies, which are the galaxies that lack
what can be interpreted as the structural signatures of
major---``wet'' or ``dry''---mergers: surface brightness profiles with
de Vaucouleurs-like or higher \citet{Sersic:63} indices, and, absent core
scouring by binary massive black holes, power-law central luminosity
cusps \citep[e.g.,][and references
  therein]{Hopkins:08,Hopkins:09a,Hopkins:09b}.  Imaging with the {\it
  Hubble Space Telescope} (HST) has revealed that $75\%$ of late-type
(Sc--Sd) disk galaxies contain compact, luminous star clusters at
their centers with masses in the range
$8\times10^5-6\times10^7\,M_\odot$, effective radii in the range
$1-9\,\textrm{pc}$ with a median of $\sim 3.5\,\textrm{pc}$, and
velocity dispersions in the range
$13-34\,\textrm{km}\,\textrm{s}^{-1}$
\citep[e.g.,][]{Boker:02,Boker:04,Walcher:05}.  The clusters' ages
range widely from $4\times10^7$ to $10^{10}\,\textrm{yr}$, the
metallicities average at $\langle Z\rangle=0.015$ with a significant
scatter, the average star formation rate over the last
$100\,\textrm{Myr}$ is $\langle\Sigma_{\rm
  SFR}\rangle=2\times10^{-3}\,M_\odot\,\textrm{yr}^{-1}$, and older
stars dominate the cluster mass.  Similarly, an HST survey of stellar
nuclei in spheroidal galaxies in the Virgo and Fornax clusters
\citep{Ferrarese:06a,Cote:06,Miller:07} has revealed stellar nuclei
that on average contain $\sim 0.2\%$ of the galaxy mass
\citep{Ferrarese:06b,Wehner:06} and have a median effective radius of
$4.2\,\textrm{pc}$ \citep[see, also,][]{Geha:02}.  Also, HST imaging
of nearby low-luminosity dwarf galaxies with absolute magnitudes
$-16<M_V<-13$ \citep{Georgiev:09} has revealed NSCs with masses $\sim
(10^6-10^7)\,M_\odot$, with a majority clustering at the lower end of
the mass range, and half-light radii $1.5-4.7\,\textrm{pc}$.

It seems that photometrically distinct NSCs are common in galaxies
that have not experienced major mergers in the epoch following the
initial burst of star formation \citep[see, e.g.,][]{Koda:09}, but
have a well-defined dynamical center. The faintest dwarf spheroidal
satellites of the Milky Way, with the exception of the Sagittarius
Dwarf \citep[e.g.,][and references therein]{Bellazzini:08}, do not
contain NSCs, even though they do seem to form a continuous structural
family of stellar systems with the more massive dwarf spheroidals that
do contain NSCs \citep[e.g.,][and references
  therein]{Kormendy:09,Wyse:10}. While the spheroidal galaxies in
Virgo are typically non-rotating \citep{Geha:03}, their
nucleus-subtracted surface brightness profiles, with S\'ersic indices
$n\sim 1-2$, resemble those of disks.  This motivates a unified
approach to addressing NSC formation in these two morphological
classes.  In disks, an excess surface brightness above the
inward-extrapolated exponential law of the outer disk occurring at
radii $\sim 100-500\,\textrm{pc}$, variously characterized as
``central light excess'' \citep{Boker:03} or a ``pseudobulge''
\citep{Kormendy:04,Fisher:08,Fisher:10,Fisher:09,Weinzirl:09}, may
provide additional clues, because the process that produced the NSC
could also contribute to the formation of the pseudobulge.
Pseudobulges can be distinguished from classical bulges through their
cold, rapidly rotating kinematics and low S\'ersic index
\citep[see][and references therein]{Kormendy:08}.  \citet{Kormendy:10}
find that the pseudobulges in the sample of pure-disk (Scd) galaxies
contain $\lesssim 3\%$ of the galactic stellar mass,
which is smaller than the stellar mass fraction in classical bulges.

Here we discuss the formation of NSCs in the context of normal star
formation in which most stars form in clusters.  A successful theory
of NSC formation should explain the observed uniformity of their
properties. The smallness of their masses in comparison with the total
stellar masses of the host galaxies is a ``small parameter''
characterizing the assembly of cosmic structure that requires
explanation.  A theory of NSC formation could help constrain the
earliest stages of the host galaxy's formation.  NSCs are interesting
because they seem to betray either an inefficiency in the transport of
baryons (gas, stars) to the center of the galaxy, or perhaps an
inefficiency of star formation once gas has arrived at the center of
the galaxy.  The NSC phenomenon may be related to the relatively slow,
or ``secular'' buildup of the host galaxy's stellar mass, since the
galaxies that form the bulk of their stars on dynamical time scales,
as in mergers \citep[e.g.,][]{Hopkins:08,Hopkins:09a} and cold
accretion from the intergalactic environment
\citep[e.g.,][]{Dekel:09a,Ceverino:10}, should turn out to be more
concentrated than the NSC host galaxies and should contain
``classical'' bulges structurally equivalent with ellipticals.

NSCs can grow from the local interstellar medium if the galactic gas
accretes and accumulates at the center \citep[e.g.,][]{Seth:06}.  In
gas disks, sufficient gas accretion to form NSCs may simply be driven
by magnetic stresses in the gas that are amplified by the
magnetorotational instability \citep{Milosavljevic:04}, which is
generic in differentially rotating galactic disks
\citep[e.g.,][]{Kim:03}.  If the inner baryonic disk is globally
self-gravitating, accretion can be driven by torques associated with
stellar and gaseous bars \citep[e.g.,][and references
  therein]{Kormendy:04,Schinnerer:07}. NSCs in late-type disks
typically contain young stellar components, and their star formation
seems to be intermittent \citep{Rossa:06,Walcher:06}, which is
consistent with the gas accretion scenario.  Cosmological
hydrodynamical simulations of dwarf and late-type galaxy formation
commonly yield nucleus- (or ``bulge''-)disk structures
\citep[e.g.,][]{Governato:04,Governato:07,Governato:09,Kaufmann:07,Brooks:09,Ceverino:10},
where the nucleus forms from the gas that self-gravitational torquing
has driven into the center of the galaxy, whereas the disk forms from
the gas that manages to stably circularize in the combined
gravitational potential of the dark matter halo and gaseous
stellar nucleus.  The resulting nuclei normally contain larger
fractions of galaxy mass than the observed nuclei in late-type disks
and spheroidals.  Frequently, a gas expulsion by feedback from star
formation is invoked to explain this apparent discrepancy
\citep[e.g.,][]{Zavala:08,Dutton:09}.

Here we do not study gas accretion, but instead investigate the
possibility, which is necessitated by the clustered character of star
formation, that a substantial fraction of the NSC mass could have
assembled in off-center stellar associations (clusters) that
subsequently migrated into the galactic centers and merged into an NSC
as intact entities
\citep[e.g.,][]{Tremaine:75,Lotz:01,Capuzzo:08a,Capuzzo:08b}.  The
assembly of NSCs through the merging of clusters is consistent with
the observation that nuclear cluster phase space densities are on
average somewhat smaller than those of globular clusters
\citep{Walcher:05} because phase space density tends to increase in
collisionless mergers. The NSCs thus assembled should inherit the
orbital angular momentum of the migrating disk clusters and should be
rotating \citep[see, e.g.,][]{Seth:08}.

\citet{Bekki:10} carried out $N$-body simulations of star cluster
orbital decay in the background of field stars in a disk galaxy
embedded in a dark matter halo.  His results suggest that NSCs could
have formed from stars delivered by inspiraling star clusters.  In a
similar spirit but different context and following an earlier proposal
by \citet{Noguchi:99}, \citet{Immeli:04}, \citet{Bournaud:07},
\citet{Elmegreen:08b}, and \citet{Ceverino:10} carried out numerical
simulations to find that the migration and central merging of giant
star clusters and gravitationally bound gas clumps in massive,
rapidly-star-forming disks can give rise to classical-bulge-like,
dynamically hot central stellar systems. If giant cluster migration
can produce classical bulges in gravitationally unstable disks that
are fed by a relatively smooth accretion from the extragalactic
environment---and not by major mergers \citep[see,
  e.g.,][]{Genzel:08}---then this warrants an investigation of cluster
migration in lower-surface-density galaxies, in hope that an improved
understanding of their morphological transformation can be gained to
elucidate why these galaxies end up lacking classical bulges and
remain rotationally-dominated and dynamically cold.

To this end, we develop a crude toy model to calculate the influx of
migrating young clusters formed in a galaxy that has experienced an
instantaneous star formation episode, but our results are relevant for
galaxies, like the prototypical late-type disk galaxy M33, in which
star formation has been steady and ongoing over the life of the
galaxy, although the timescale for the accumulation of NSC mass will
differ from the one estimated here because of the dependence, in this
case, on the star formation rate in the disk.  We assume that all stars form in clusters
and that the clusters eventually get disrupted in the galactic tidal
field.  We further discuss the assumptions and limitations of our
model in Section \ref{subsec:limitations}.  Gas expulsion, stellar
mass loss, dynamical evaporation, and external perturbations lead to
either rapid or progressive tidal mass loss in the clusters
\citep[e.g.,][]{Fall:01,Baumgardt:03,McMillan:03,Gieles:06a}.  If a
cluster is massive, it can migrate radially inward prior to complete
disruption.  For example, we will find that for our galactic models and parameters (see
Sections \ref{sec:input} and \ref{sec:results}) a $10^4\,M_\odot$
cluster never migrates more than $10\,\textrm{pc}$; a
$10^5\,M_\odot$ cluster can migrate and reach the center from
$\sim\textrm{few}\times10\,\textrm{pc}$, a $10^6\,M_\odot$ cluster can
do so from
$\sim100\,\textrm{pc}$ and a $10^7\,M_\odot$ cluster can reach the
center from
$\gtrsim1\,\textrm{kpc}$. The cluster's stars are then deposited at
galactocentric radii traversed by the migrating cluster.  Since even
in very late-type disks (Hubble type Scd) circular test particle
orbits rotate differentially throughout, cluster disruption is
inevitable except if the cluster forms at, or migrates into, the
galactic center, where tidal shear preserves the symmetries of an
axisymmetric cluster.  Clusters with off-center birthplaces can avoid
complete disruption if they migrate to the dynamical center quickly
enough.  The progressive tidal disruption of migrating clusters before
they have reached the galactic center transports mass radially inward;
we propose that this process is a driver of pseudobulge growth in
disks.

% SUMMARIZE UPCOMING SECTIONS OF PAPER
This work is organized as follows.  In Section \ref{sec:general} we
provide an empirically and theoretically motivated description of
cluster formation, migration, and dissolution.  In Section
\ref{sec:input} we introduce our models for a spheroidal galaxy and a
disk galaxy initially lacking an NSC, and estimate the effects of
galactic stellar mass redistribution due to cluster migration.  In
Section \ref{sec:results} we show our results for representative
galaxy models.  In Section \ref{sec:discussion} we compare our results
with observations of NSCs and pseudobulges and provide a brief
discussion of implications for the pseudobulge and classical bulge
dichotomy and for massive black holes, and in Section
\ref{sec:conclusions} we summarize our main conclusions.

\section{Cluster Formation, Migration and Dissolution}
\label{sec:general}
% INTRO TO CLUSTER SECTION
We assume that all stars in our model galaxies formed in clusters, and
introduce a framework based on previous empirical and theoretical work
for treating the formation, migration and dissolution of these young
stellar clusters. In Section \ref{subsec:migration} we discuss a
theoretically motivated description of cluster migration in spheroidal
and disk galaxies, taking into account the dominant form of dynamical
friction torque acting on a migrating cluster in each type of galaxy.
In Section \ref{subsec:disruption} we describe an empirically
motivated model of cluster disruption based on observations of young
clusters in disk galaxies.  In Section \ref{subsec:icmf} we introduce
the initial cluster mass function, which seems to be generic to star
formation in all galaxies, including the Milky Way.

\subsection{The Initial Cluster Mass Function}
\label{subsec:icmf}

The initial cluster mass function (ICMF) contains about equal mass on
all cluster mass scales, i.e.,
\beq
\label{eq:icmf}
\frac{dn}{dM}\propto M^{-\alpha} \ \ \ \textrm{for} \ \ \  M_{\rm min} < M < M_{\rm max},
\eeq 
with $\alpha\sim 2$
\citep[e.g.,][]{Bik:03,deGrijs:03,Hunter:03,Rafelski:05,Chandar:10},
where $M$ denotes initial cluster mass and $M_{\rm min}$ and $M_{\rm
  max}$ are galaxy-dependent cutoffs. We adopt $\alpha=2$ and $M_{\rm
  min}=100\,M_\odot$ in what follows; our results will depend only
weakly on $M_{\rm min}$ since clusters with masses below
$\sim10^4\,M_\odot$ are disrupted before they can migrate to the center
from larger radii and, if they do reach the center, contribute an
insignificant fraction of the total NSC mass.  The definition of
a high-mass cutoff of the ICMF and the statistical significance of
observational evidence for such a cutoff have been the subject of
debate.  We assume here that the ICMF is indeed subject to high mass
truncation, and treat $M_{\rm max}$ as a free parameter that varies
over the range $10^4\,M_\odot-10^7\,M_\odot$, which includes the
cutoffs reported for the nearby galaxies. We will find in Section
\ref{sec:results} that masses of nuclear clusters in our calculations
are most sensitive to the truncation mass, and thus the photometry, of
spheroidal and disk galaxies. If combined with a theoretical model of
cluster migration and dissolution, this sensitivity can be utilized to
indirectly constrain the ICMF truncation mass scale in these galaxies.

The maximum cluster mass forming in NGC 6946, M51, and the Antennae
has been estimated from observations to be $M_{\rm
  max}\sim10^6\,M_\odot$ \citep{Gieles:06b,Gieles:06c}.  Frequently, an
ICMF with an exponential cutoff $dn/dM\propto
M^{-\alpha}e^{-M/M_\star}$ is found to adequately approximate the
truncation of the ICMF at the high-mass end. In spirals and irregulars
the mass truncation scales $M_\star$ and $M_{\rm max}$ are in the
range $\sim 2\times 10^5 - 10^{7.5} M_\odot$ \citep[see,
  e.g.,][]{Zhang:99,Bik:03,deGrijs:03,Dowell:08,Larsen:09,Gieles:09}.
The exponential mass truncation scale, which has typical values of
$M_\star\sim 2\times 10^5 M_\odot$, is larger in denser starburst
environments \citep{Larsen:09}.  In the present work, we adopt the
abrupt truncation with $M_{\rm max}$, but we do not anticipate that
the results would be substantially different if exponential truncation
were used.

The maximum mass of clusters that formed in the assembly of nucleated
spheroidals is unknown, but a brief theoretical speculation may be in
order.  In a gas disk with surface density $\Sigma_{\rm gas}$ close to
the critical value $\sim 10\, M_\odot\,\textrm{pc}^{-2}$ corresponding
to the column density $N\sim 10^{21}\,\textrm{cm}^{-2}$ that is
required for the presence of a self-shielding cold phase \citep[see,
  e.g.,][]{Schaye:04} and Toomre parameters $Q\lesssim 1$, the
truncation scale $M_{\rm max}$ (or $M_\star$) should scale with the
Jeans mass of the disk,
\bea
\label{eq:Jeans_mass}
M_{\rm  max}&\sim& f_{\rm SF} \,\left(\frac{4\pi^2G\Sigma_{\rm
      gas}}{\kappa^2}\right)^2\Sigma_{\rm gas}\nonumber\\
&\sim& 4\times10^6\,M_\odot\,f_{\rm SF,-1}\,\left(\frac{\Sigma_{\rm
      gas}}{10\,M_\odot\,\textrm{pc}^{-2}}\right)^3
\left(\frac{\kappa}{10^{-15}\,\textrm{s}^{-1}}\right)^{-4} ,
\eea
where $f_{\rm SF} =0.1 f_{\rm SF,-1}$ is the star formation efficiency
and $\kappa$ is the epicyclic frequency \citep[see,
  e.g.,][]{Elmegreen:08b}.  The reference epicyclic frequency in
equation (\ref{eq:Jeans_mass}) was selected to correspond to the
characteristic average total mass density $\sim
0.1\,M_\odot\,\textrm{pc}^{-3}$ in the inner $300\,\textrm{pc}$ of the
(largely non-nucleated) dwarf spheroidal satellites of the Milky Way
\citep{Strigari:08}, but of course, the proto-spheroidals would have
been characterized by larger, radius-dependent values of $\kappa$.  If
the gas disk mass is a fixed proportion of the total (e.g., dark
matter) enclosed mass, and the vertical scale height of the disk is
$h$, then $\kappa\propto \Sigma_{\rm gas}^{1/2}\,h^{-1/2}$, which
would imply a weaker dependence $M_{\rm max}\propto \Sigma_{\rm gas}
\, h^2$ on the gas surface density.  It is plausible that
proto-spheroidals assembled from an ICMF reaching the cluster mass
scale associated with the gas disk Jeans mass estimated in equation
(\ref{eq:Jeans_mass}).

\subsection{Cluster Migration}
\label{subsec:migration}

A gravitationally-bound star cluster migrates on a time scale 
\beq \label{eq:time_mig}
t_{\rm mig}= \frac{dJ}{d\ln R}\left<T_{\rm DF}\right>^{-1} ,
\eeq
where $R$ is a characteristic size of the orbit---the equivalent of
the semimajor axis in a non-Keplerian potential, $J$ is the angular
momentum of the cluster, and $\left<T_{\rm DF}\right>$ is the
orbit-averaged dynamical friction torque.  All galactic mass
components (dark matter, gas disk, stars) respond dynamically to the
cluster but the mechanism of torque coupling varies.  We separately
consider the dynamical friction torque in spheroidal and disky mass
components.

In spheroidal stellar systems or dark matter halos, $N$-body
simulations have shown that the torque from a nonrotating
collisionless halo can be heuristically described with the
Chandrasekhar formula for dynamical friction,
\beq
\label{eq:torque_spheroid}
\vec{T}_{\rm DF,Chandra}=\frac{4\pi\ln(\Lambda) \chi(V) G^2 M^2 \rho }{V^3}\,\vec{r}\times\vec{V}, 
\eeq
where $\rho$ is the local combined density of stars and dark matter,
$|\vec{V}|\sim \Omega R$ is the velocity of the cluster,
$\ln(\Lambda)$ is the Coulomb logarithm, and $\chi(V)$ is the mass
fraction of stars or dark matter particles with velocities less than
$V$.  Because the kinematic structure of actual halos differs from the
premises of Chandrasekhar's derivation, $N$-body simulations are
necessary to obtain the correct normalization of the torque amplitude.
The numerically evaluated torques
\citep[e.g.,][]{Velazquez:99,Penarrubia:02,Penarrubia:04,Spinnato:03}
can be modeled with Chandrasekhar's formula if the Coulomb logarithm
is treated as an empirically-calibrated free parameter. The values of
the Coulomb logarithm obtained from these $N$-body simulations are
$\ln(\Lambda)\sim 2-7$.

In differentially rotating gas or stellar disks, the torque is
provided by the flow in the corotation region and by angular momentum
transfer at Lindblad resonances \citep[e.g.,][]{Goldreich:80,Quinn:86}.
Disk clusters with masses $M\lesssim 10^6\,M_\odot$ have Roche tidal
radii $r_{\rm t}\sim [GM/(d\Omega/d\ln R)^2]^{1/3}$ that are smaller than the
thickness of the disk. In this regime, the disk torque is a
generalization of the ``Type I'' torque acting on small planets in
protoplanetary disks \citep[e.g.,][]{Tanaka:02,DAngelo:03,Baruteau:08}
to non-Keplerian disks and is proportional to
\beq
\label{eq:torque_disk}
T_{\rm DF,disk}\propto \frac{G^2 M^2\Sigma}{\sigma^2},
\eeq
where $\Sigma$ is the surface density of the gas or stellar disk, and
$\sigma$ is the gas sound speed or the stellar velocity dispersion in
the disk.  In what follows we take the constant of proportionality in
equation (\ref{eq:torque_disk}) to be equal to unity.  This choice seems
consistent with the numerical calibration of the torque in a
collisionless particle disk by \citet{Donner:93} and \citet{Wahde:96}
if the gravitational softening employed in their calculations is
equated with the vertical thickness of the disk.

Finally, we note that \citet{Milosavljevic:04} argued against the
migratory scenario in disks on the grounds of long migration
times. There, migration from a distant location in the disk was
envisioned, and the treatment ignored the enhancement in dynamical
friction force due to the stellar accumulation of earlier migrating
clusters in the central kiloparsec of the galaxy.  In the present
picture, clusters migrate from a range of radii in the galactic disk,
but only those that form closest to the galactic center reach the
center and merge with the nuclear cluster.

\subsection{Cluster Dissolution}
\label{subsec:disruption}

The timescale on which a cluster is disrupted in the galactic tidal
field has been determined empirically by modeling the luminosity and
age functions of clusters in nearby disk galaxies \citep[see,
  e.g.,][]{Lamers:05a,Lamers:05b}.  Theoretical models tracking
stellar and dynamical evolution of a cluster in a tidal field
\citep[e.g.,][]{Gieles:08a, Gieles:08b} have reproduced the
observationally inferred variation of the dissolution time with
cluster mass,
\beq
\label{eq:time_dis}
t_{\rm dis}= t_0\left(\frac{M}{M_\odot}\right)^\gamma ,
\eeq
where $\gamma\approx 0.62$ \citep{Boutloukos:03,deGrijs:06}. The
normalization $t_0$ varies between galaxies, which seems to be a
consequence of the variation in the tidal field strength and of
cluster-scale density inhomogeneities of the galactic environment.
\citet{Gieles:08a} assessed the role of the tidal field by comparing
cluster lifetimes in several galaxies with the inverse angular
frequency of circular orbits at observed cluster radii and found
consistency with the linear relation $t_{\rm dis}\propto
\Omega^{-1}$. The residual variation of $t_{\rm dis}$ between galaxies
has been ascribed to disruption by giant molecular clouds \citep[][and
  references therein]{Gieles:06a} which we ignore.  The tidal radius
really depends on the degree of differential rotation, $r_{\rm
  t}=[GM/(d\Omega/d\ln R)^2]^{1/3}$ \citep[e.g.,][]{Quinn:86},
%$r_{\rm t}=|GM/(d\Omega^2/d\ln R)|^{1/3}$ \citep[e.g.,][]{Binney:08},
and thus one would expect that $t_0\propto|d\Omega/d\ln R|^{-1}$.
\citet{Lamers:05a} modeled the observed cluster population in M33 and
found, at an arbitrary radius, $t_0\approx f_{\rm dis}/\Omega(R)$ with
$f_{\rm dis,M33}\approx 0.16$. 

The coefficient $f_{\rm dis}$, which
can be separately inferred from the observed young cluster populations and from theoretical calculations,
encapsulates the detailed mass loss mechanics in a cluster 
embedded in a galactic tidal field and
evolving through internal two-body relaxation.
From the theoretical viewpoint, the coefficient
depends on the initial stellar mass function, on the initial
cluster structure, and on the (possibly time-dependent) gravitational
potential within and near the cluster.  Under idealized assumptions,
the coefficient has been estimated with $N$-body
simulations \citep{PortegiesZwart:98,PortegiesZwart:02,Baumgardt:03}.
The theoretical estimate $f_{\rm dis}\sim 0.3$ in
\citet{Lamers:05a}, which is based on the results of \citet{Baumgardt:03}, is larger than the
empirical value.  The clusters originating in M33's central disk
could be denser and more resistant to tidal disruption than those in
the sample of \citet{Lamers:05a}.

The above relations were obtained for clusters
originating in a galactic disk. For lack of an equivalent
empirical result for spheroidal starbursts, we assume that these
relations hold universally and adopt the crude relation
\beq
\label{eq:t_zero}
t_0 = f_{\rm dis}\left|\frac{d\Omega}{d\ln r}\right|^{-1},
\eeq
where we take $f_{\rm dis} = 0.2$. We have compared our simplified
form of the dissolution timescale defined in equations
(\ref{eq:time_dis}) and (\ref{eq:t_zero}) for a cluster with initial
mass $M=10^5\,M_\odot$ and angular velocity $\Omega(r)=[G M_{\rm
    gal,0}(r)/r^3]^{1/2}$, where $M_{\rm
  gal,0}(r)=M_{\star,0}(r)+M_{\rm DM}(r)$ is the total initial
galactic mass within a sphere of radius $r$ (see Section
\ref{sec:input}), with the timescale from
\citet{Lamers:05a,Lamers:05b}, where $t'_{\rm
  dis}=t'_4(M/10^4\,M_\odot)^\gamma$ \citep{Boutloukos:03}, $t'_4=1.355
\,\textrm{ yr}\times 10^{4 \gamma}\gamma^{-1}(t'_0/1\,\textrm{
  yr})^{0.967}$ \citep{Lamers:05b}, and $t'_0 = f_{\rm
  dis}\Omega(r)^{-1}$, and found it to be consistent to within $20\%$
in the range of radii where we find that significant cluster migration
takes place.

\section{Initial and Redistributed Galactic Stellar Mass}
\label{sec:input}

Here we describe our models for the initial density profile of a
spheroidal galaxy (Section \ref{subsec:spheroidal}) and a disk galaxy
(Section \ref{subsec:disk}). The initial density lacks a stellar
nucleus; its innermost baryonic density profile is an extrapolation
from larger radii. Then in Section \ref{subsec:redistribution}, we
describe our method of distributing initial clusters consistent with
these surface density profiles and modeling cluster mass
redistribution during migration and dissolution. We include the
growing central cluster mass in the mass distribution affecting
subsequent migrating clusters, but we ignore the response of the dark
matter halo to the baryonic collapse and subsequent mass
redistribution by cluster migration.

\subsection{Spheroidal Galaxy Model}
\label{subsec:spheroidal}

Photometry of spheroidal galaxies in the nearby universe
\citep{Cote:06,Ferrarese:06a,Kormendy:09}, combined with the
fact that in these relatively old stellar systems the stellar mass-to-light ratios do not vary
significantly within galaxies, has shown that their mass surface
density profiles are well described by a two-component model
\citep{Balcells:03,Graham:03}
\beq
\label{eq:stellar_profile}
\Sigma_\star(R) = \Sigma_{\rm Sersic} (R) + \Sigma_{\rm nucl}(R) .
\eeq
The S\'ersic law component 
$
\Sigma_{\rm Sersic}(R) \propto \exp[-(R/R_{\rm s})^{1/n}]
$
\citep{Caon:93},where $n$ is the S\'ersic index and $R_{\rm s}$ is a scale
radius, contains most of the mass, and the compact nuclear component
$\Sigma_{\rm nucl}(R)$ contains a small fraction
\citep{Ferrarese:06b,Wehner:06} of the mass.  In spheroidals the
S\'ersic index varies in the range $n\sim 1-2$
\citep{Ferrarese:06a,Kormendy:09}.  We work within the paradigm in
which a stellar nucleus is not present prior to the formation of most
of the galactic stars. Thus, we consider an initial galaxy with a
spherically-averaged stellar density profile that lacks a nucleus.  
We employ an approximation to the deprojected S\'ersic profile in the
form \citep{Prugniel:97,LimaNeto:99}
\beq
\rho_{\star,0}(r) = \rho_{\rm sph}\left(\frac{r}{R_{\rm s}}\right)^{-p}\exp\left[-\left(\frac{r}{R_{\rm s}}\right)^{1/n}\right],
\eeq
where 
\beq
\label{eq:p_of_n}
p = 1 -  0.6097 \, n^{-1} + 0.05463 \, n^{-2} .
\eeq 
Here, $\rho_{\rm sph}$ is normalized to the total stellar mass of the galaxy
$M_{\rm sph}=4\pi\int_0^\infty\rho_{\star,0}(r)r^2dr=4\pi n R_{\rm s}^3
\rho_{\rm sph}\Gamma[(3-p)n]$, and $\Gamma(a)=\int_0^\infty
t^{a-1}e^{-t}dt$ is the gamma function.  Quantities pertaining to the
initial galaxy, before migration and disruption of clusters, are
denoted by the subscript $0$.  The stellar mass within radius $r$ is
\beq
M_{\star,0}(r) = 4\pi n R_{\rm s}^3 \rho_{\rm sph} \gamma[(3 - p)n, (r/R_{\rm s})^{1/n}],
\eeq
where $\gamma(a,x)=\int_0^x t^{a-1}e^{-t}dt$ is
the lower incomplete Gamma function.

The total initial density profile includes a dark matter component
\beq
\label{eq:total_profile}
\rho_0(r) = \rho_{\star,0}(r) + \rho_{\rm DM}(r) , 
\eeq
where we model the dark matter density profile with the NFW law
\citep{Navarro:97} $\rho_{\rm DM}(r) = \rho_s/[(r/r_s)(1 + r/r_s)^2]$
with scale radius $r_{\rm s}$, virial radius $r_{\rm vir}$, and
concentration $c\equiv r_{\rm vir}/r_{\rm s}$.  Spheroidals are often
dark matter dominated, especially at the low-luminosity end, i.e.,
$\rho_{\rm DM}(r)\gtrsim \rho_{\star,0}(r)$. Thus, for galaxies in
which dark matter is the dominant mass component, the cluster
migration time can be modeled with the $N$-body-simulation-calibrated
Chandrasekhar dynamical friction torque in equation
(\ref{eq:torque_spheroid}).

\subsection{Disk Galaxy Model}
\label{subsec:disk}

The surface density profiles of late-type spiral disk galaxies can be
fit with a S\'ersic-type exponential stellar disk and a nuclear
component \citep{Boker:03} as in equation
(\ref{eq:stellar_profile}). We consider ``pure'' disk galaxies, which
do not have a stellar bulge or a preexisting pseudobulge; we do not
consider cluster formation in the galactic halo.  The initial stellar
density profile without the nuclear component is then
\beq 
\Sigma_{\star,0}(R) =
\Sigma_{\rm disk}\exp\left[-\left(\frac{R}{R_{\rm s}}\right)^{1/n}\right] , 
\eeq 
where $\Sigma_{\rm disk} =M_{\rm disk} /
[2\pi n R_{\rm s}^2 \Gamma(2n)]$. The corresponding initial
stellar mass profile is
\beq
M_{\star,0}(R) = 2\pi n R_{\rm s}^2\Sigma_{\rm disk} \Gamma[2n,(R/R_{\rm s})^{1/n}],  
\eeq 
where $\Gamma(a,x)=\Gamma(a)-\gamma(a,x)$ is the upper incomplete
Gamma function.
The total initial density profile is then
\beq 
\Sigma_0(R) = \Sigma_{\star,0}(R) + \Sigma_{\rm DM}(R), 
\eeq 
where $\Sigma_{\rm DM}(R)$ is the projected NFW profile. 
%via the Abel transform. 
Disk galaxies generally have an exponential profile with S\'ersic
index $n\sim1$.

The total density profiles in disk galaxies may be dominated either by
dark matter or by luminous matter in their central regions. Even if
dark matter dominates the spherically averaged density profile, in
disk galaxies it provides a relatively smaller contribution to the
torque if the cluster is embedded in the disk. This is because the
torque coupling of the migrating star cluster with the flattened disk,
equation (\ref{eq:torque_disk}), is more significant than that with
the dark matter halo \citep[see, e.g.,][]{Bekki:10}. We use the disk
torque to model the migration time in disk galaxies.

\subsection{Stellar Mass Redistribution by Cluster Migration}
\label{subsec:redistribution}

We assume that all stars form in clusters with masses between
$M_{\rm min}$ and $M_{\rm max}$ (see Section \ref{subsec:icmf}). Let
$d^2n/dMdr$ be the number of clusters per unit initial cluster mass
per unit radius that have formed in a galaxy such that $\int_0^\infty\,
(d^2n/dMdr)\,dr$ is proportional to the ICMF in equation
(\ref{eq:icmf}), while
\beq
\int_{M_{\rm min}}^{M_{\rm max}} \frac{d^2n}{dMdr}\,MdM = 
\begin{cases}
4\pi r^2\rho_{\star,0}(r), & \textrm{spheroid}, \cr
         2\pi r\Sigma_{\star,0}(r), & \textrm{disk} .
\end{cases}
\eeq
We sample cluster masses and initial radii randomly according to the
distribution $d^2n/dMdr$.  The total number of clusters in our simulations is
$
\int_{M_{\rm min}}^{M_{\rm max}} (dn/dM)\,dM \approx M_{\rm tot} / [M_{\rm min}\ln(M_{\rm max}/M_{\rm min})], 
$
where $M_{\rm tot}$ is the total stellar mass, $M_{\rm sph}$ or
$M_{\rm disk}$, of the galaxy.  At first, we do not explicitly take into account
the possible prompt, possibly mass-independent \citep[for a discussion
  of mass-dependent prompt dissolution see, e.g.,][]{Goodwin:09}
dissolution of clusters \citep[``infant mortality,'' see,
  e.g.,][]{Fall:05,Bastian:06,Chandar:06,Goodwin:06,Gieles:07,Pellerin:07,Whitmore:07,Baumgardt:08,deGrijs:08,Lamers:09}.
Then, we briefly assess the effects of prompt dissolution.

We proceed to model cluster migration and dissolution by following the
cluster orbital decay
\beq
\frac{dr}{dt} = - \frac{r}{t_{\rm mig}} ,
\eeq 
and mass loss 
\beq
\frac{dM}{dt} = -\frac{M}{t_{\rm dis}} ,
\eeq
where $t_{\rm mig}$ and $t_{\rm dis}$ are, respectively, the migration
time scale in equation (\ref{eq:time_mig}) and dissolution time scale
in equation (\ref{eq:time_dis}).

We calculate the orbital decay of clusters in the order of increasing
migration time (see equation [\ref{eq:time_mig}]) evaluated before
migration has occurred.  We keep track of the mass tidally stripped
from the cluster with $k$th shortest migration time by calculating its
contribution to the stellar density profile
\beq
\Delta \rho_{\star,k} = \frac{1}{4\pi r_k^2} \frac{dM_k}{dt}
\left/\frac{dr_k}{dt}\right. \ \ \  (r_k< r_{k,0}) ,
\eeq
where $r_{k,0}$ is the cluster formation radius and $k=1,2,\ldots
n_{\rm tot}$.  We assume that a cluster has been fully disrupted when
its mass falls below $M_{\rm min}=100\,M_\odot$ since in clusters of
this size
$t_{\rm mig}\gg t_{\rm dis}$ and thus the
cluster is not able to migrate substantially, if at all, before
disruption; this residual mass is deposited at the radius of
disruption.  Some migrating clusters avoid complete disruption until
they reach the innermost radius of our grid, $r_{\rm
  min}=1\,\textrm{pc}$.  The migrating cluster mass that reaches the
central parsec is added to the central point mass $M_{\rm cent}$.  To
account for the influence of the growing nuclear component, including
$M_{\rm cent}$ and the mass deposited at larger radii, e.g., at
$r\lesssim100\,\textrm{ pc}$, on the migration and disruption time
scales of more slowly inspiraling clusters, we calculate the tidally
stripped mass of faster migrating star clusters with smaller $t_{\rm
  mig}$ first.  After the migration and disruption of $k$ clusters has
been computed, we sum up their contributions to the modified stellar
density profile of the galaxy $\Delta \rho_\star^{(k)} = \sum_{k'=1}^k
\Delta \rho_{\star,k'}$, and set the stellar density profile in which
the $(k+1)$st cluster migrates to $ {\rm max}(\rho_{\star,0},\Delta
\rho_\star^{(k)})$ so that the late-migrating clusters do so in a
galaxy modified by the central concentration increase from the
early-migrating clusters.  We take the final density profile of the
galaxy, after all clusters have been disrupted, to be
$\rho_\star=\sum_k\Delta \rho_{\star,k}$.  This prescription does not
account for the escape of the material liberated by stellar mass loss
from the galaxy.  To model the effect of this escape, we repeat the
calculation after requiring that the first $50\%$ of the mass to be
stripped from a cluster leave the galaxy.

We treat the mass delivered to the central parsec as an ``unresolved''
stellar nucleus component with mass $M_{\rm cent}$.  We model its surface
density profile with the function
\beq
\Sigma_{\rm cent}(R) = \frac{M_{\rm cent}\, R_{\rm cent}^2}{\pi (R_{\rm
 cent}^2+R^2)^2} , 
\eeq
where $R_{\rm cent}$ is the half-light radius of the unresolved component. We set this radius to $R_{\rm
  cent}=2\,\textrm{ pc}$, which is chosen to be larger than our
innermost grid radius but smaller than typical observed half-light
radii of NSCs, e.g.,
$R_{\rm nuc}\gtrsim10\,\textrm{pc}$ \citep[e.g.,][]{Geha:02}.  We add this unresolved component to
the surface density profile calculated from the mass deposited by
migrating clusters at $r>1\,\textrm{pc}$, so that
\beq
\Sigma_{\star,{\rm total}} (R) = \int_{-\infty}^\infty
\rho_\star[(R^2+z^2)^{1/2}]dz  + \Sigma_{\rm cent} (R) .
\eeq
Our surface density profiles do not take into account any smearing by
the point spread function (PSF).  If the PSF width is comparable to
the resulting NSC radius, then smearing by the PSF must be taken into
account when comparing synthesized surface brightness profiles with
the observed ones.  Because our results are affected by a variety of
crude approximations we do not proceed to model the effect of the
PSF.
 
\subsection{Limitations of the Model}
\label{subsec:limitations}

Before proceeding to discuss our results we would like to highlight
the assumptions and limitations of our model.  We have assumed that
all clusters form simultaneously and thus we calculated the orbital decay
of the clusters in order from fastest-migrating (as calculated before any migration has occurred) to
slowest-migrating.  This assumption is applicable to those dwarf galaxies
that consume gas and form stars in a single star formation episode.
In other galaxies, including late-type disks, however,
cluster formation is ongoing (e.g., in the galaxy M33), or is triggered by galactic
mergers (e.g., in the Antennae galaxies).  Our model could be adapted 
to approximate ongoing star formation by calculating cluster
orbital decay and the associated mass transport in random order so
that an older, less massive, and slower-migrating cluster
can migrate and deposit its mass before a younger, more massive, and faster-migrating
cluster does.  Without migration-time ordering, our resulting NSC
masses are similar to those in the corresponding
ordered models.  We do not attempt to estimate the time scale for the
build-up of NSC mass in this
scenario because the 
total timescale, although certainly longer than in the instantaneous
star formation scenario, would depend on additional parameters---the
star formation rate and the rate of gas accretion onto the
galaxy---that are bound to vary between galaxies.  

In an effort to focus on the variation in the galactic surface density
profile and NSC mass buildup as a function of the ICMF truncation mass
scale $M_{\rm max}$, we do not study their dependence on other
parameters such as galaxy mass and initial concentration (S\'ersic
index), or on parameters, such as the
coefficients characterizing the amplitudes of the dynamical friction
torques driving cluster migration, that are subject to theoretical
uncertainty.  
In the same spirit, we do not 
consider star formation episodes resulting from galactic mergers as
this would require separate tracking of the galactic stellar and gas
masses.  The role of gaseous accretion and galactic mergers in
galaxy assembly is best addressed with comprehensive cosmological
hydrodynamical simulations
\citep[e.g.,][]{Governato:04,Governato:07,Governato:09,Kaufmann:07,Brooks:09,Ceverino:10},
as analytical and semi-analytical models are inadequate in this context.
We also do not model the response of the dark matter halo to
the initial baryonic infall and to the subsequent evolution due to
young cluster migration and stellar mass loss. Our model for progressive cluster migration
and decay is crude and thus we will compare only the general
characteristics of the resulting surface density profiles with those
of observed galactic profiles. We caution against comparison with
the detailed features of observed galaxy photometry.

\section{Results}
\label{sec:results}

We are ready to define the parameters of our sample calculations and
provide results for model spheroidal and disk galaxies forming at high
and low redshifts.  For a fixed effective radius of the initial
stellar density profile, the formation redshift determines the
average density of dark matter within the effective
radius. High-redshift models have dark matter halos with lower
concentration and smaller virial radius than low-redshift models.  The
parameters $M_{\rm halo}$, $c$, and $r_{\rm vir}$, in general, depend
on the assumed cosmological model and on the redshift of halo
formation. We assume a flat universe with $\Omega_{\rm m}=0.274$,
$\Omega_\Lambda=0.726$, and $h=0.705$ \citep{Komatsu:09}. The mass of
the halo is related to the virial radius by $M_{\rm halo} =
\frac{4}{3}\pi\Delta_{\rm c}(z)\rho_{\rm crit}r_{\rm vir}^3$, where
$\rho_{\rm crit}$ and $\Delta_{\rm c}(z)$ are, respectively, the critical
density of the universe and the mean overdensity within the virial
radius at redshift $z$
%$\approx 18\pi^2 - 82\Omega_\Lambda - 39\Omega_\Lambda^2
%\approx 100$ is the mean density at redshift $z=0$.
\citep{Bryan:98}.  For example, $\Delta_{\rm c}(0)\approx100$ and
$\Delta_{\rm c}(5-10)\approx177$.  The median concentration parameter
for all high-redshift dark matter halos is $c\approx3.5$, but at low
redshift varies steeply with halo mass
\citep[e.g.,][]{Bullock:01,Zhao:03,Gao:08,Klypin:10}.  Our high- and
low-redshift models are for $z=6$ and $z=0$, respectively.  

We present our results for spheroidal and disk galaxies in Sections
\ref{subsec:sph_results} and \ref{subsec:disk_results}, respectively.
Then in Section \ref{subsec:prompt} we explore the
observationally-motivated scenario in which a large fraction of the
clusters dissolve promptly following formation. 

\subsection{Spheroidal Galaxies}
\label{subsec:sph_results}

\begin{figure*}
\begin{center}
\includegraphics[width=0.3\textwidth,clip]{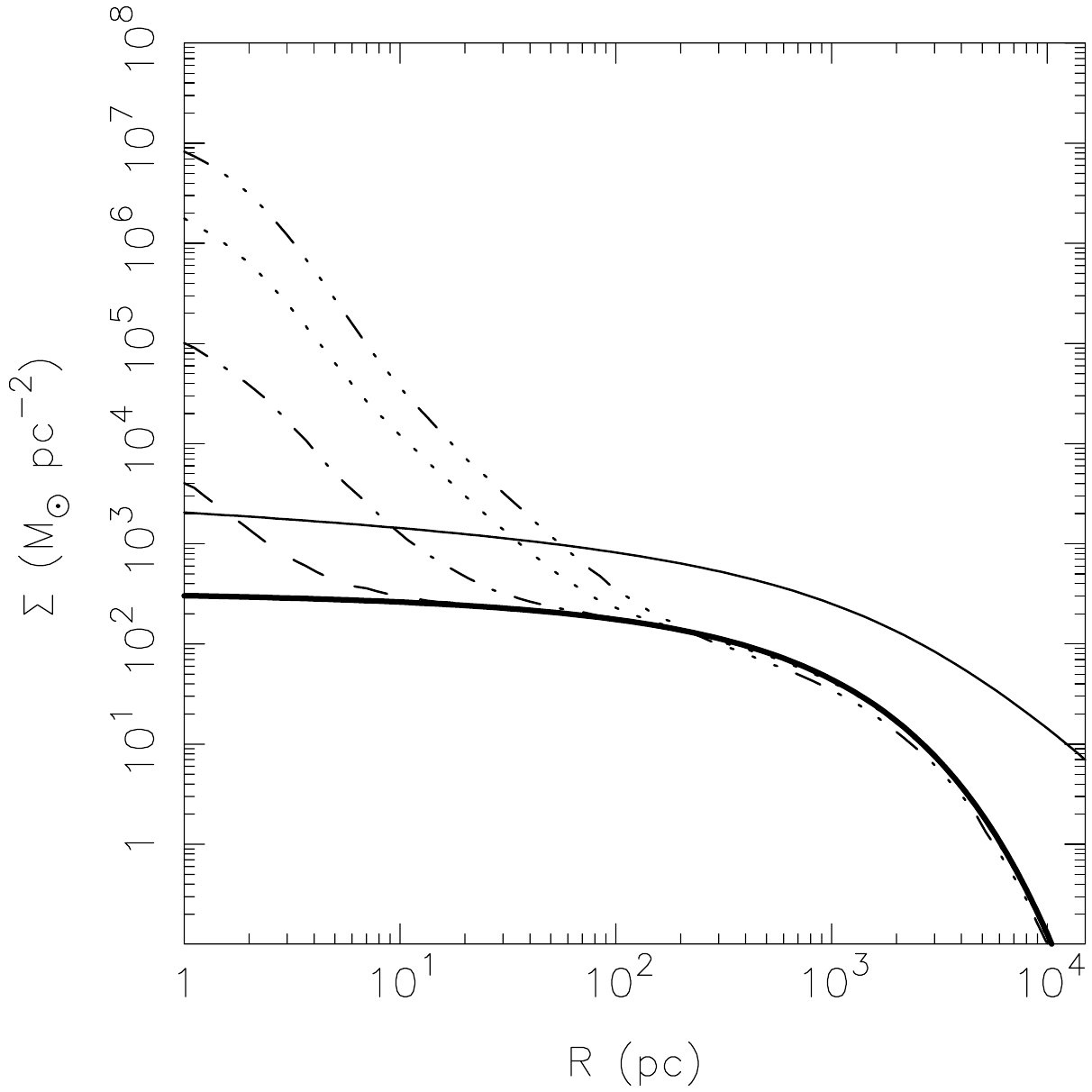}
\includegraphics[width=0.3\textwidth,clip]{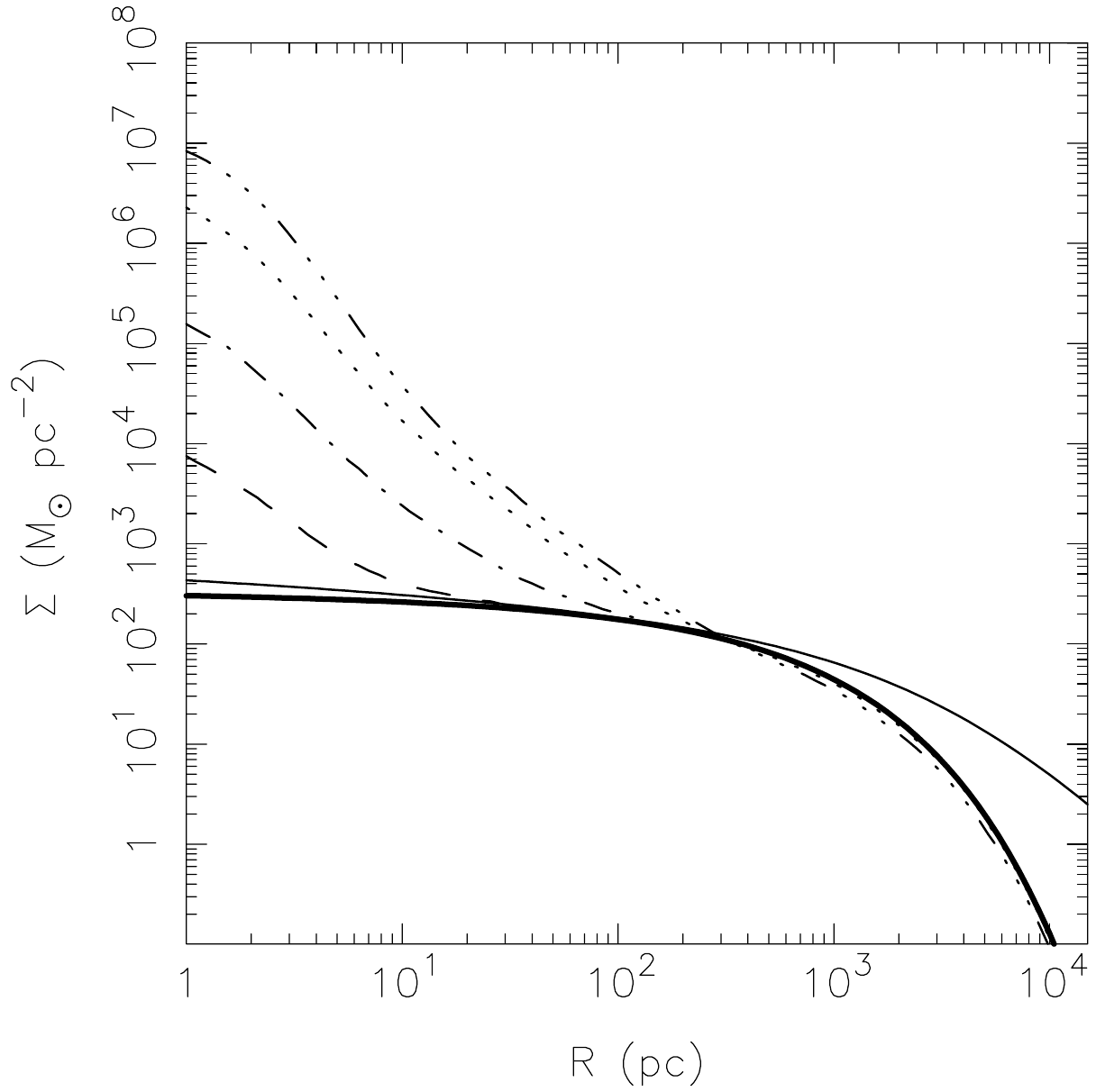}
\includegraphics[width=0.3\textwidth,clip]{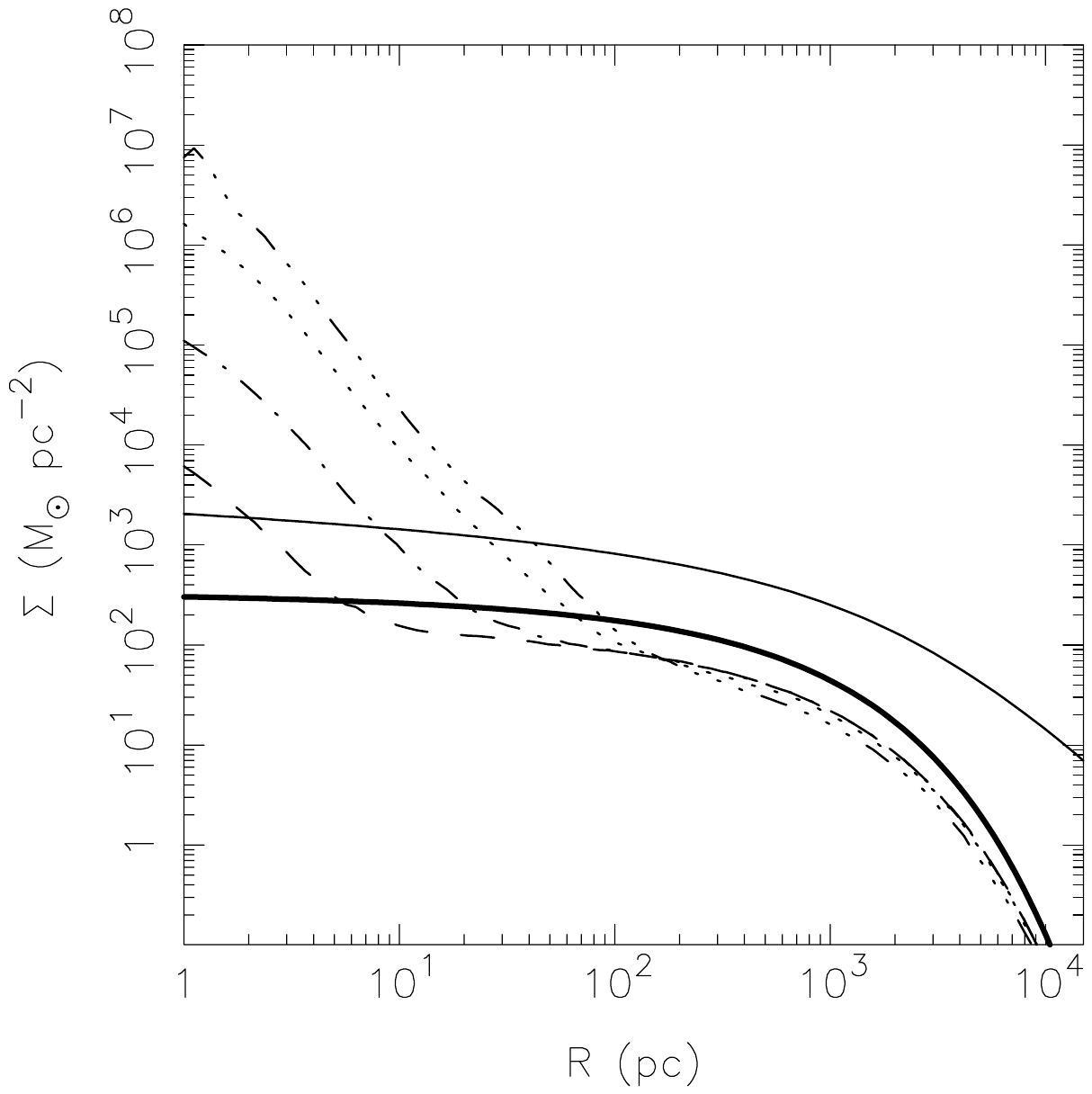}
\end{center}
\caption{Projected initial surface density profiles of stars (thick
  solid line) and dark matter (thin solid line) for a proto-spheroidal
  galaxy and the final stellar surface density profile of the galaxy
  after all the migrating clusters have been disrupted, for ICMF
  high-mass truncation of $M_{\rm max}=10^4 M_\odot$ (dashed line) ,
  $10^5 M_\odot$ (dash-dotted line), $10^6 M_\odot$ (dotted line),
  $10^7 M_\odot$ (dash-triple-dotted line; see Section
  \ref{subsec:icmf}), and the effective Coulomb logarithm
  $\ln\Lambda=5$ (see Section \ref{subsec:migration}). We show
  spheroidal galaxies formed at high redshift (left panel) and at low
  redshift (middle panel) ignoring galactic mass loss due to stellar
  evolution, and assuming $50\%$ mass loss (right panel; see Section
  \ref{subsec:redistribution}).}
\label{fig:profiles_sph}
\end{figure*}

In Figure \ref{fig:profiles_sph}, left and middle panel, we show the
projected surface density profiles for stars before migration and
disruption ($M_{\rm sph}=10^{9}\,M_\odot$, $R_{\rm s}=0.5\,\textrm{kpc}$,
$n=1.5$), for dark matter ($M_{\rm halo}=10^{10}\,M_\odot$, high-z:
$c=3.5$, $r_{\rm vir}=10\,\textrm{ kpc}$; low-z: $c=15$, $r_{\rm
  vir}=60\,\textrm{ kpc}$), and for stars after migration and
disruption, assuming $\ln\Lambda=5$ and $M_{\rm
  max}=(10^4,\,10^5,\,10^6,10^7)\,M_\odot$.  To simplify our
calculation, we take $p= 2/3$ instead of the value $p\approx 0.62$
implied by equation (\ref{eq:p_of_n}).  In these models we have not
taken into account galactic mass loss resulting from stellar
evolution.  To explore the consequences of potential galactic loss of
gas liberated by supernovae and AGB stars, in Figure
\ref{fig:profiles_sph}, right panel, we repeat the calculation for the
high-redshift model assuming that the first $50\%$ of the mass
dissociated from each migrating cluster is lost, i.e., leaves the
galaxy.  The models shown in Figure \ref{fig:profiles_sph} do not
allow for prompt cluster dissolution; we explore the impact of prompt
dissolution in Section \ref{subsec:prompt} below.

The final structure of the galaxy, after all clusters have migrated
and dissolved, is a strong function of the ICMF truncation mass scale
$M_{\rm max}$.  For $M_{\rm max}=10^4\,M_\odot$, the excess surface
density above the initial stellar density profile is present only in
the central $\sim 10\,\textrm{pc}$ and is small.  At the other end of
the range of ICMF truncation mass scales, for $M_{\rm
  max}=10^7\,M_\odot$, the surface density profile of the entire
galaxy has become slightly more concentrated, but still approximately
follows the initial S\'ersic profile at large radii. For $M_{\rm
  max}=(10^4-10^7)\,M_{\rm max}$, the resulting surface density excess
after cluster migration and disruption appears as a distinct departure
from the inward-extrapolated outer S\'ersic law at
$R\sim20-200\,\textrm{pc}$ in the high-$z$ model and
$R\sim30-300\,\textrm{pc}$ in the low-$z$ model; the excess surface
density $\Sigma_{\star,\rm{total}}-\Sigma_{\rm cent}$ approximates
power laws $\propto R^{-0.6}-R^{-1.7}$.  There is little difference
between the profiles calculated at low and high redshift; this could
in part be attributed to the cancellation of the nearly opposite
effects of increasing halo concentration and decreasing mean halo
density with decreasing redshift.  Galactic stellar mass loss from
stellar evolution leads to a decrease of stellar surface density at
all radii but does not modify the overall character of the final
surface density profile.

\begin{figure*}
\begin{center}
\includegraphics[width=0.45\textwidth,clip]{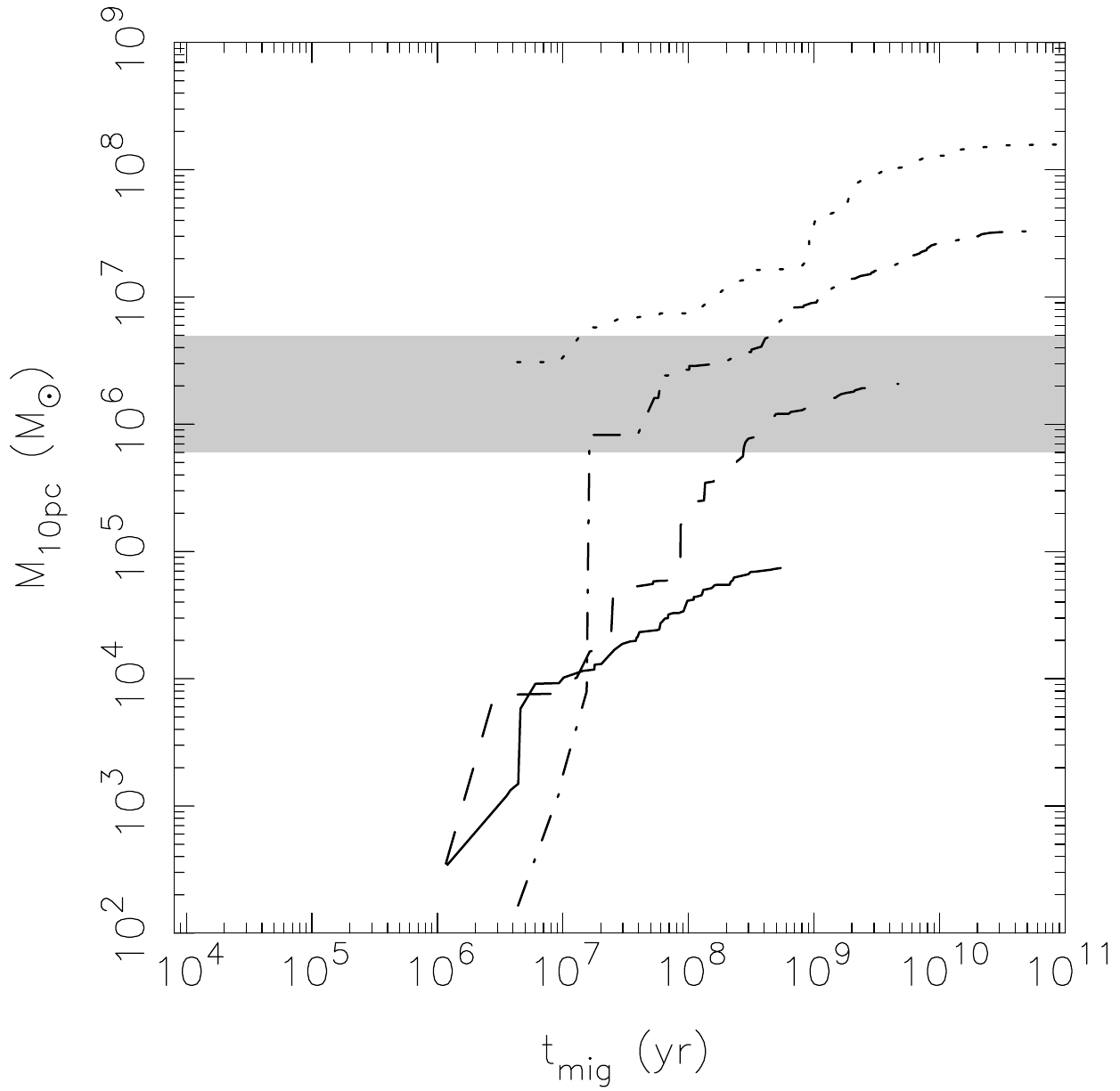}
\includegraphics[width=0.45\textwidth,clip]{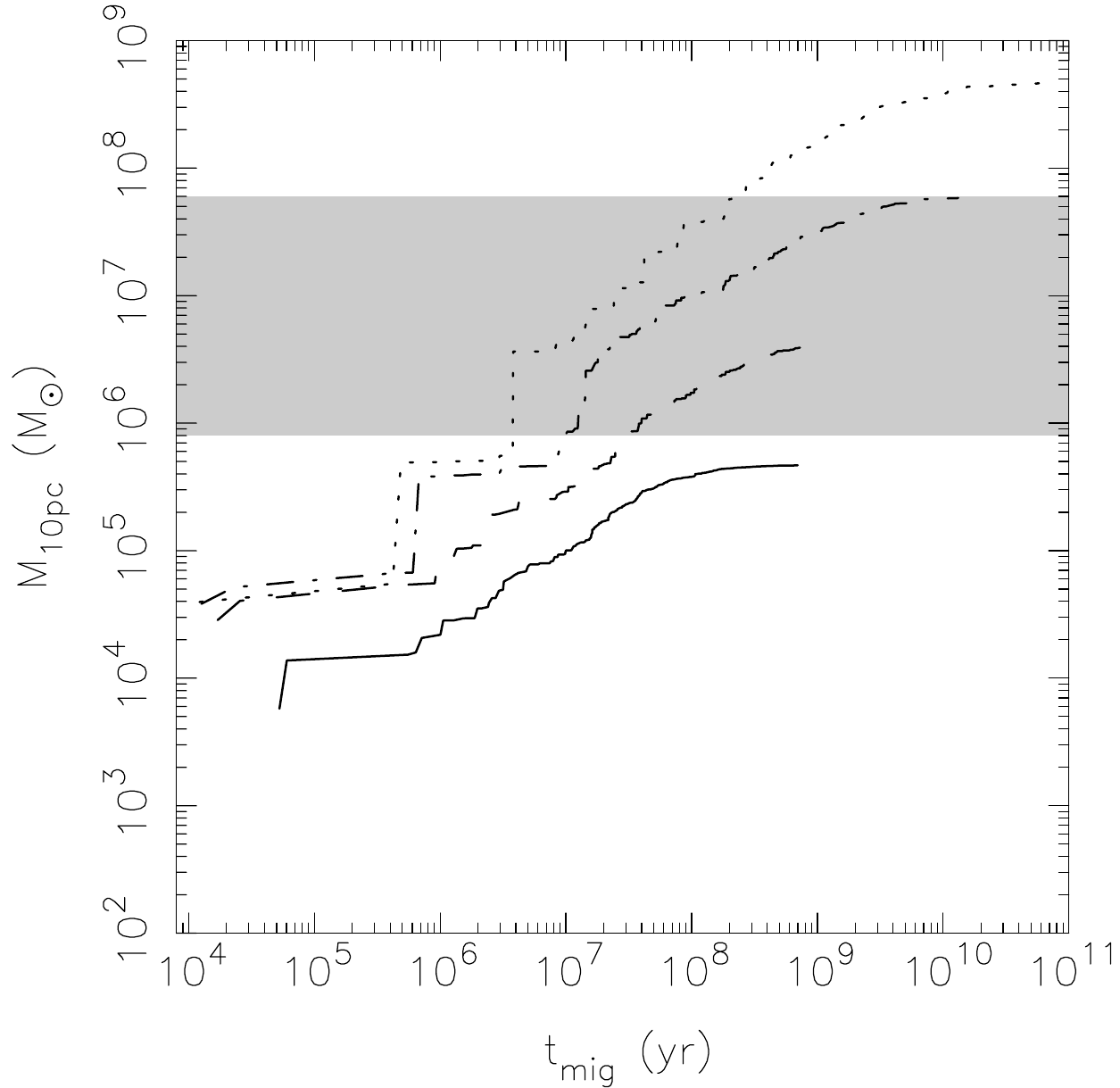}
\end{center}
\caption{Total mass $M_{10\,{\rm pc}}$ accumulated after migration and
  disruption of star clusters within the central $10\,\textrm{pc}$ of
  a high-redshift spheroidal galaxy (left panel) and disk galaxy
  (right panel) as a function of migration time for those clusters
  that reach the central $10\,\textrm{pc}$ for $M_{\rm max}=10^4
  M_\odot$ (solid line), $M_{\rm max}=10^5 M_\odot$ (dashed line),
  $10^6 M_\odot$ (dash-dotted line), and $10^7 M_\odot$ (dotted line).
  The mass histories end when the last cluster makes it to the central
  $10\,\textrm{pc}$.  The shaded region in the left panel shows the
  $1\sigma$-range of NSC mass-to-total stellar mass ratios from
  \citet{Ferrarese:06b} and \citet{Cote:06}.  In the right panel, the
  shaded region shows the range of NSC masses in \citet{Walcher:05}.}
\label{fig:mass_time_high_z}
\end{figure*}

In Figure \ref{fig:mass_time_high_z}, left panel, we show the
evolution of the NSC mass proxy $M_{10\,{\rm pc}}$, here simply
calculated as the stellar mass contained in a sphere with radius
$r=10\,\textrm{pc}$ as a function of the initial migration time of the
$k$th cluster to migrate to the galaxy center (see Section
\ref{subsec:redistribution}).  The latter is an approximation since
the migration time evolves as the cluster loses mass and as its orbit
decays, and as the galactic potential is modified by the previous
$k-1$ migrating clusters.  The initial value of $M_{10\,\rm{pc}}$ in
Figure \ref{fig:mass_time_high_z}, ranging from $\sim10^2\,M_\odot$ to
$>10^6\,M_\odot$, corresponds to the initial mass of the fastest-migrating cluster in each
simulation; for $M_{\rm max}=10^7\,M_\odot$, the mass of the
fastest-migrating cluster to reach $r<10\,\textrm{pc}$ already falls
within the range of observed NSC masses (see Section
\ref{subsec:observations}).  The mass $M_{10\,{\rm pc}}$
  contains $96\%$ of the mass of the unresolved nucleus $M_{\rm cent}$
  accounting for the cluster stars that have reached
  $r=1\,\textrm{pc}$ and an additional ``resolved'' mass deposited by
  clusters migrating from $10\,\textrm{pc}$ to $1\,\textrm{pc}$.  The
mass $M_{10\,{\rm pc}}$ increases as an approximate power-law in time
with slopes between $M_{10\,{\rm pc}}\propto t^{1/2}$ and $M_{10\,{\rm
    pc}}\propto t$ and then levels off.  In Table \ref{tab:sph} we
report the final asymptotic $M_{10\,{\rm pc}}$ and its ratio to the
total stellar mass, as well as the total time $t_{1/2}$ that it takes
for $M_{10\,{\rm pc}}$ to reach one-half of its final value.  In all
the models presented in this work, the contribution of the unresolved
central point mass $M_{\rm cent}$ to $M_{10\,{\rm pc}}$ increases from
about $40\%$ for $M_{\rm max}=10^4\,M_\odot$ to $86\%$ for $M_{\rm
  max}=10^7\,M_\odot$ in spheroidals and from about $20\%$ for $M_{\rm
  max}=10^4\,M_\odot$ to $80\%$ for $M_{\rm max}=10^7\,M_\odot$ in
disks.  We find that $>90\%$ of the NSC mass is composed of clusters
with initial masses $>0.1\,M_{\rm max}$, indicating that the
lowest-mass clusters make up only a small fraction of the NSC mass
and, thus, that the final central mass is not sensitive to $M_{\rm
  min}$.

The asymptotic $M_{10\,{\rm pc}}$ is a strong function of the ICMF
truncation mass scale $M_{\rm max}$, as is the time scale for approach
to this asymptotic mass.  Larger $M_{\rm max}$ yield larger final
central masses that are assembled on longer time scales; this is
because more massive clusters are able to migrate from larger radii.
The central mass-to-ICMF truncation mass ratios are $M_{10\,{\rm
    pc}}/M_{\rm max}\sim 1-1.5\,\textrm{dex}$, and thus the central
stellar system is assembled from many migrating clusters.  The scaling
of the resulting NSC mass with the ICMF characteristic mass scale is
consistent with the trend recognized by \citet{Bekki:10}, who carried
out $N$-body integrations with an initial lognormal cluster mass
function that approximates the evolved present-day globular cluster
mass distribution.

\subsection{Disk Galaxies}
\label{subsec:disk_results}

\begin{figure*}
\begin{center}
\includegraphics[width=0.3\textwidth,clip]{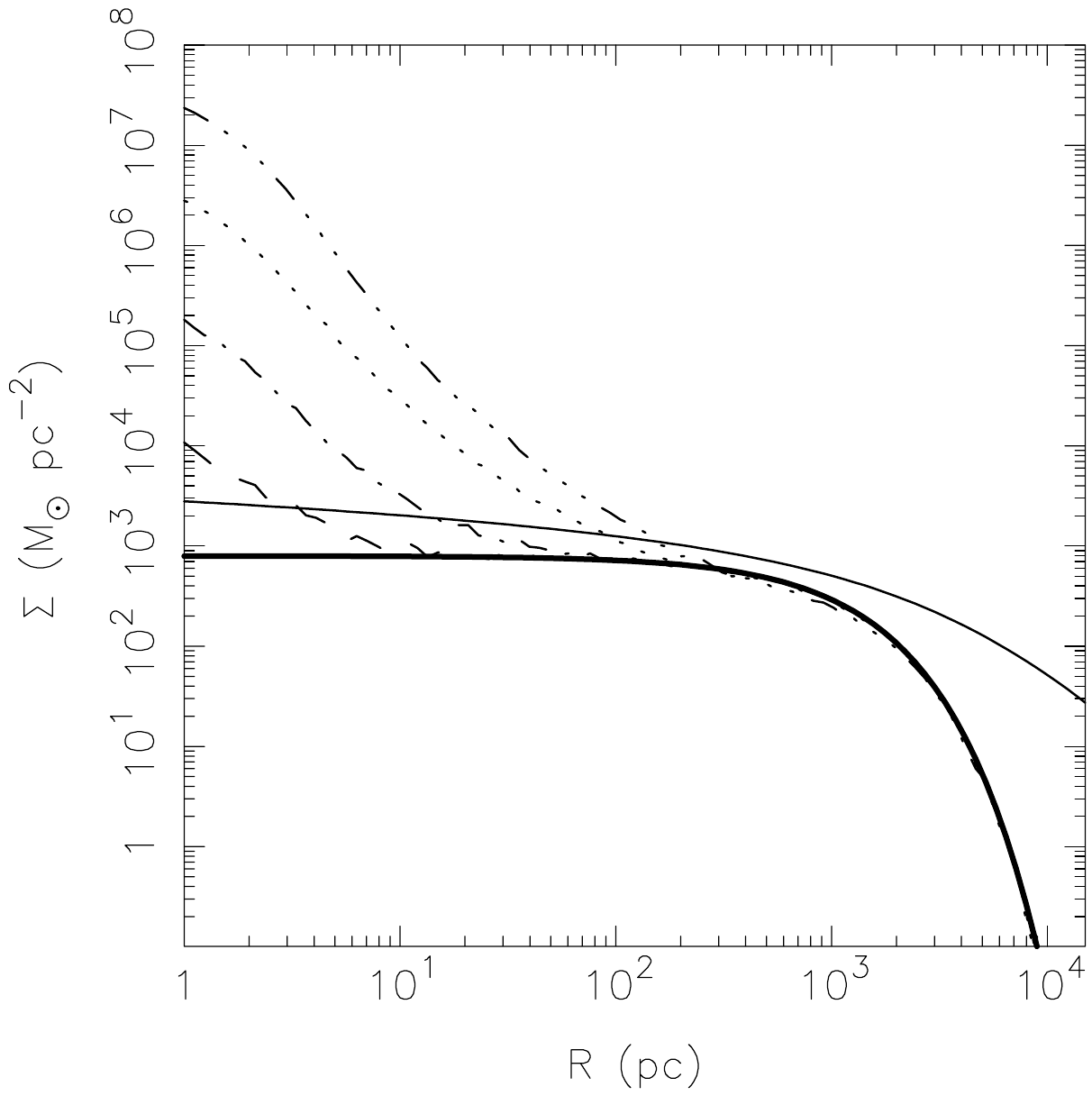}
\includegraphics[width=0.3\textwidth,clip]{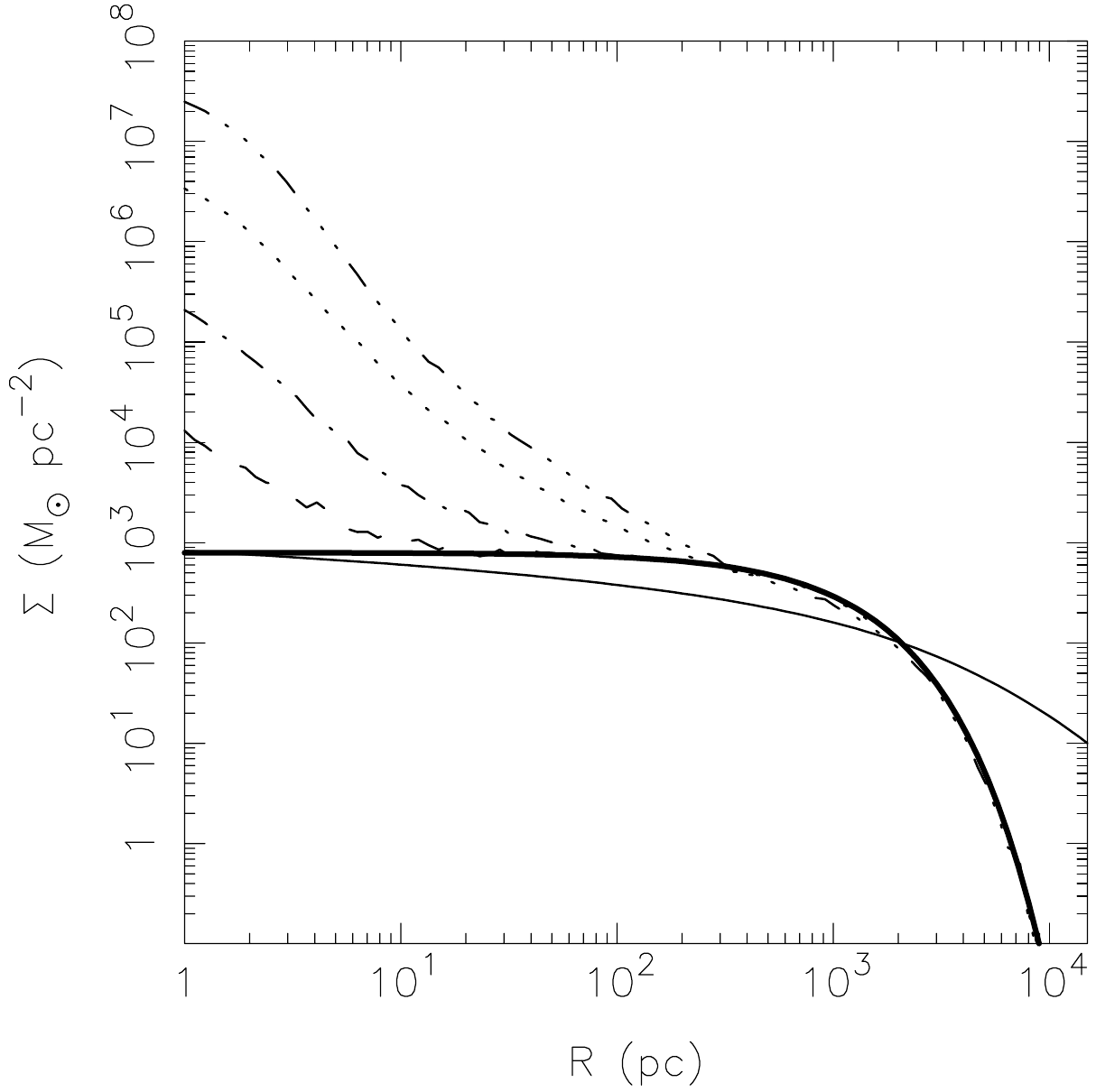}
\includegraphics[width=0.3\textwidth,clip]{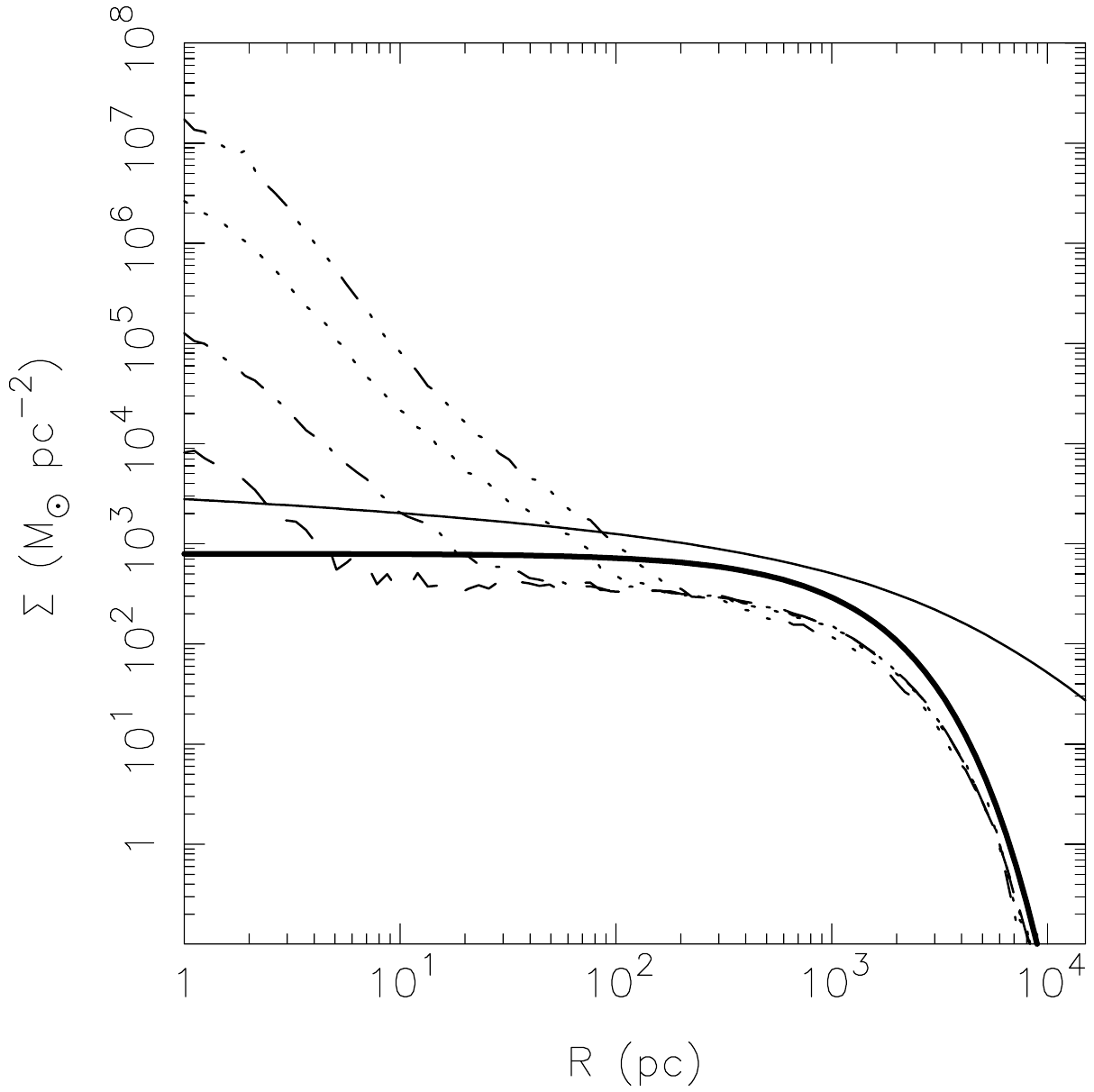}
\end{center}
\caption{Same as Figure \ref{fig:profiles_sph}, but for our disk
  galaxy model at high redshift (left panel) and low redshift (middle
  panel), and at high redshift with $50\%$ stellar mass loss (right
  panel).}
\label{fig:profiles_disk}
\end{figure*}

In the disk galaxy calculation we choose $n=1$ and assume an
approximate gas or stellar velocity dispersion in equation
(\ref{eq:torque_disk}) of $\sigma(r) = 0.2\,[GM_{\rm gal}(r)/r]^{1/2}$
where $M_{\rm gal}(r)={\rm max} [M_{\star,0}(r),M_{\star,k}(r)]+M_{\rm
  DM}(r) + M_{{\rm cent},k}(r)$ (see Section
\ref{subsec:redistribution}). Figure \ref{fig:profiles_disk} in its
left and middle panels shows the projected surface density profiles
for stars ($M_{\rm disk}=5\times10^{9}\,M_\odot$,
$R_{\rm s}=1.5\,\textrm{kpc}$, $n=1$), dark matter ($M_{\rm
  halo}=5\times10^{10}\,M_\odot$, high-z: $c=3.5$, $r_{\rm
  vir}=20\,\textrm{ kpc}$; low-z: $c=15$, $r_{\rm
  vir}=100\,\textrm{kpc}$), and stars after migration and disruption
for $M_{\rm max}=(10^4,\,10^5,\,10^6,\,10^7)\,M_\odot$.  As we did for
spheroidal galaxies in Section \ref{subsec:sph_results}, here we test
the effect of galactic mass loss due to stellar evolution in Figure
\ref{fig:profiles_disk}, right panel. The character of a disk galaxy's
morphological transformation in consequence of cluster migration is
very similar to that observed in spheroidal galaxies. The densities at
smaller radii $\lesssim 10\,\textrm{pc}$ in disk galaxy models are
somewhat higher than in spheroidal galaxy models.  This may be
explained by a higher initial central stellar surface density in our
disk models.  In Figure \ref{fig:mass_time_high_z}, right panel, we
show the evolution of the mass enclosed within $r=10\,\textrm{pc}$ and
in Table \ref{tab:disk} we show the final, asymptotic mass
$M_{10\,{\rm pc}}$ and the time scale $t_{1/2}$ for accumulation of a
half the final mass in disk galaxy models.  We find that $M_{10\,{\rm
    pc}}/M_{\rm max}\sim 1.7\,\textrm{dex}$ and that $> 90\%$ of the
NSC mass is composed of clusters with initial masses above $0.1\,M_{\rm
  max}$ for $M_{\rm max}>10^4\,M_\odot$ indicating that, again, the
central stellar system is assembled from many migrating clusters and
that the NSC mass is not sensitive to $M_{\rm min}$.

\begin{deluxetable}{cccccc}
\tablecolumns{6}
\tablecaption{Spheroidal Galaxy\tablenotemark{a} NSC Properties}
\tablehead{\colhead{Model}
 & \colhead{$\log(M_{\rm max})$}
 & \colhead{$\log(t_{1/2})$}
 & \colhead{$\log(M_{10\,{\rm pc}})$}
 & \colhead{$M_{10\,{\rm pc}}/M_{\rm sph}$}
 & \colhead{$M_{10\,{\rm pc}}/M_{\rm max}$} \\
 & \colhead{$(M_\odot)$}
 & \colhead{(yr)}
 & \colhead{$(M_\odot)$}
 & \colhead{(dex)}
 & \colhead{(dex)} }
\startdata
\multirow{4}{*}{High-z}
& $4$ & $7.96$ & $4.87$ & $-4.13$ & $0.87$\\
& $5$ & $8.60$ & $6.32$ & $-2.68$ & $1.32$\\
& $6$ & $9.50$ & $7.52$ & $-1.48$ & $1.52$\\
& $7$ & $9.34$ & $8.20$ & $-0.80$ & $1.20$\\
\hline
\multirow{4}{*}{Low-z}
& $4$ & $8.09$ & $5.24$ & $-3.76$ & $1.24$\\
& $5$ & $8.83$ & $6.51$ & $-2.49$ & $1.51$\\
& $6$ & $9.59$ & $7.64$ & $-1.36$ & $1.64$\\
& $7$ & $9.13$ & $8.21$ & $-0.79$ & $1.21$
\enddata
\tablenotetext{a}{$M_{\rm sph}=10^{9}\,M_\odot$, $M_{\rm
    halo}=10^{10}\,M_\odot$, no galactic mass loss or prompt dissolution.}
\label{tab:sph}
\end{deluxetable}

\begin{deluxetable}{cccccc}
\tablecolumns{6}
\tablecaption{Disk Galaxy\tablenotemark{a} NSC Properties}
\tablehead{\colhead{Model}
 & \colhead{$\log(M_{\rm max})$}
 & \colhead{$\log(t_{1/2})$}
 & \colhead{$\log(M_{10\,{\rm pc}})$}
 & \colhead{$M_{10\,{\rm pc}}/M_{\rm disk}$}
 & \colhead{$M_{10\,{\rm pc}}/M_{\rm max}$} \\
 & \colhead{$(M_\odot)$}
 & \colhead{(yr)}
 & \colhead{$(M_\odot)$}
 & \colhead{(dex)}
 & \colhead{(dex)} }
\startdata
\multirow{4}{*}{High-z}
& $4$ & $7.49$ & $5.67$ & $-4.03$ & $1.67$\\
& $5$ & $8.08$ & $6.60$ & $-3.10$ & $1.60$\\
& $6$ & $8.87$ & $7.77$ & $-1.93$ & $1.77$\\
& $7$ & $9.29$ & $8.67$ & $-1.03$ & $1.67$\\
\hline
\multirow{4}{*}{Low-z}
& $4$ & $7.44$ & $5.75$ & $-3.95$ & $1.75$\\
& $5$ & $8.01$ & $6.68$ & $-3.02$ & $1.68$\\
& $6$ & $8.92$ & $7.86$ & $-1.84$ & $1.86$\\
& $7$ & $9.10$ & $8.69$ & $-1.00$ & $1.69$
\enddata
\tablenotetext{a}{$M_{\rm disk}=5\times10^{9}\,M_\odot$, $M_{\rm
    halo}=5\times10^{10}\,M_\odot$, no galactic mass loss or prompt dissolution.}
\label{tab:disk}
\end{deluxetable}

\subsection{Prompt Cluster Dissolution}
\label{subsec:prompt}

\begin{figure*}
\begin{center}
\includegraphics[width=0.3\textwidth,clip]{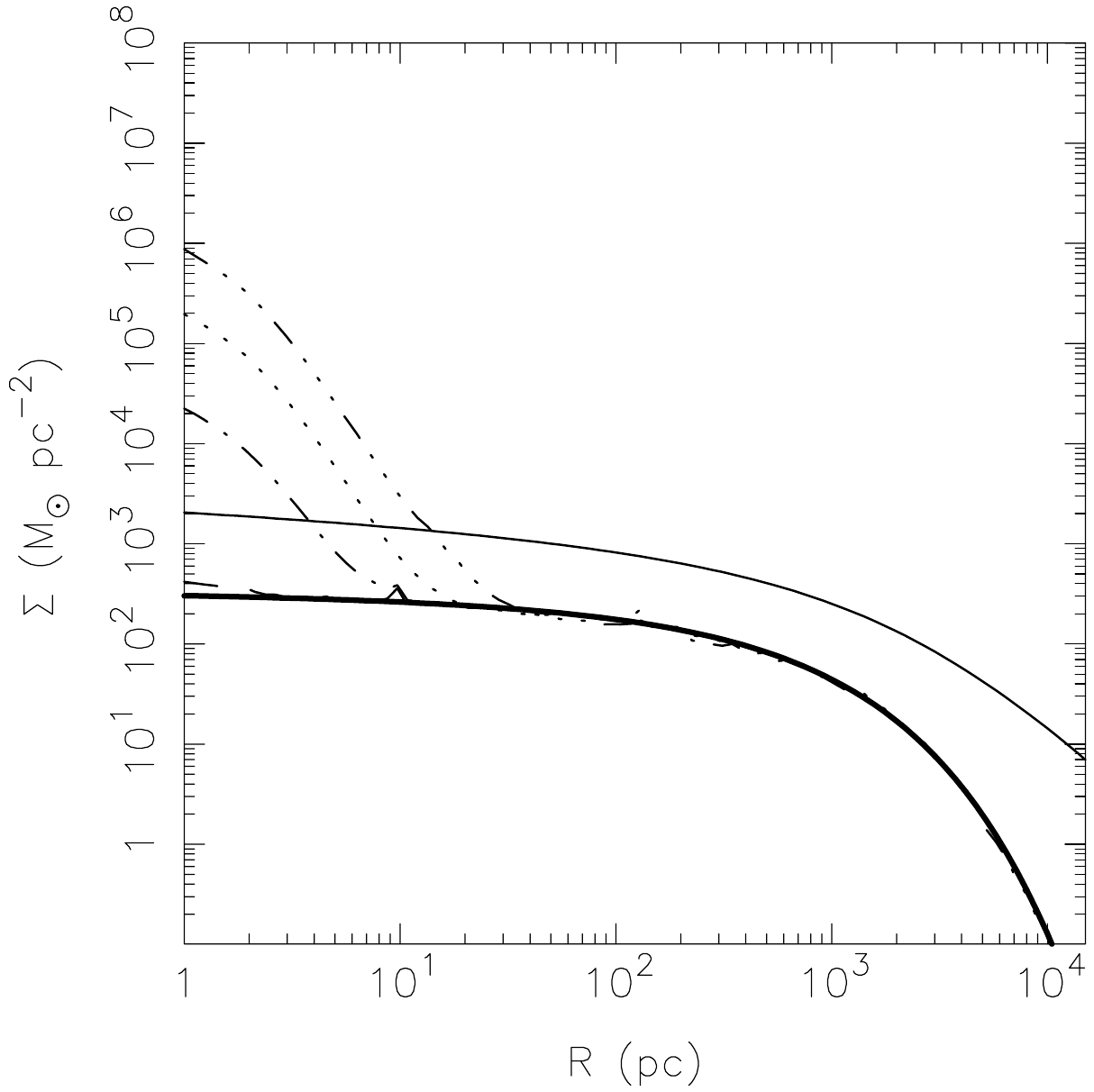}
\includegraphics[width=0.3\textwidth,clip]{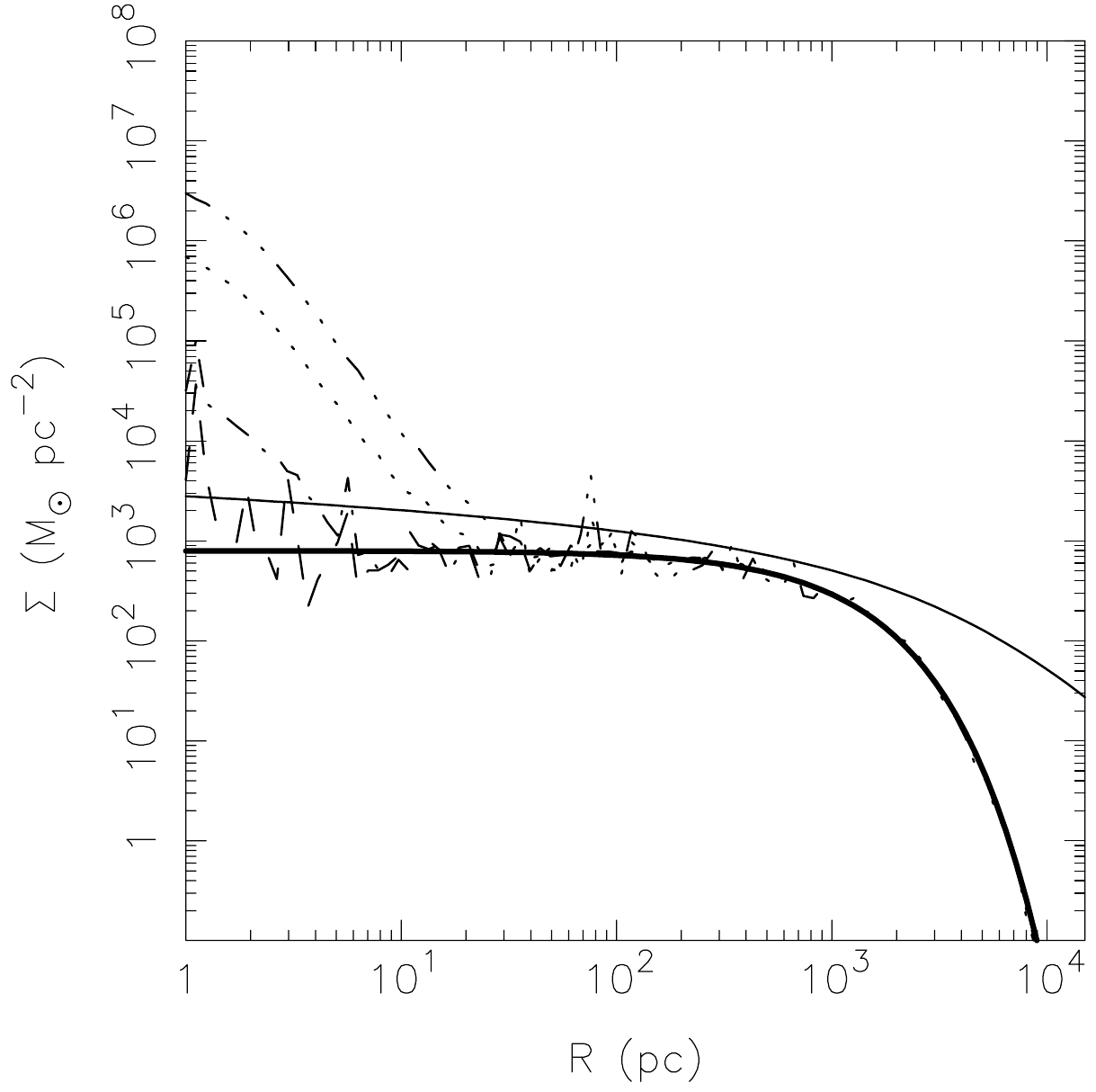}
\end{center}
\caption{Same as Figure \ref{fig:profiles_sph}, but for a high-redshift
 proto-spheroidal galaxy (left panel) and proto-disk galaxy (right
 panel) with prompt cluster dissolution.
In these models $90\%$ of the clusters dissolve
immediately independent of their mass, and the remaining $10\%$
migrate inward prior to disruption.}
\label{fig:profiles_high_z_prompt}
\end{figure*}

\begin{figure*}
\begin{center}
\includegraphics[width=0.45\textwidth,clip]{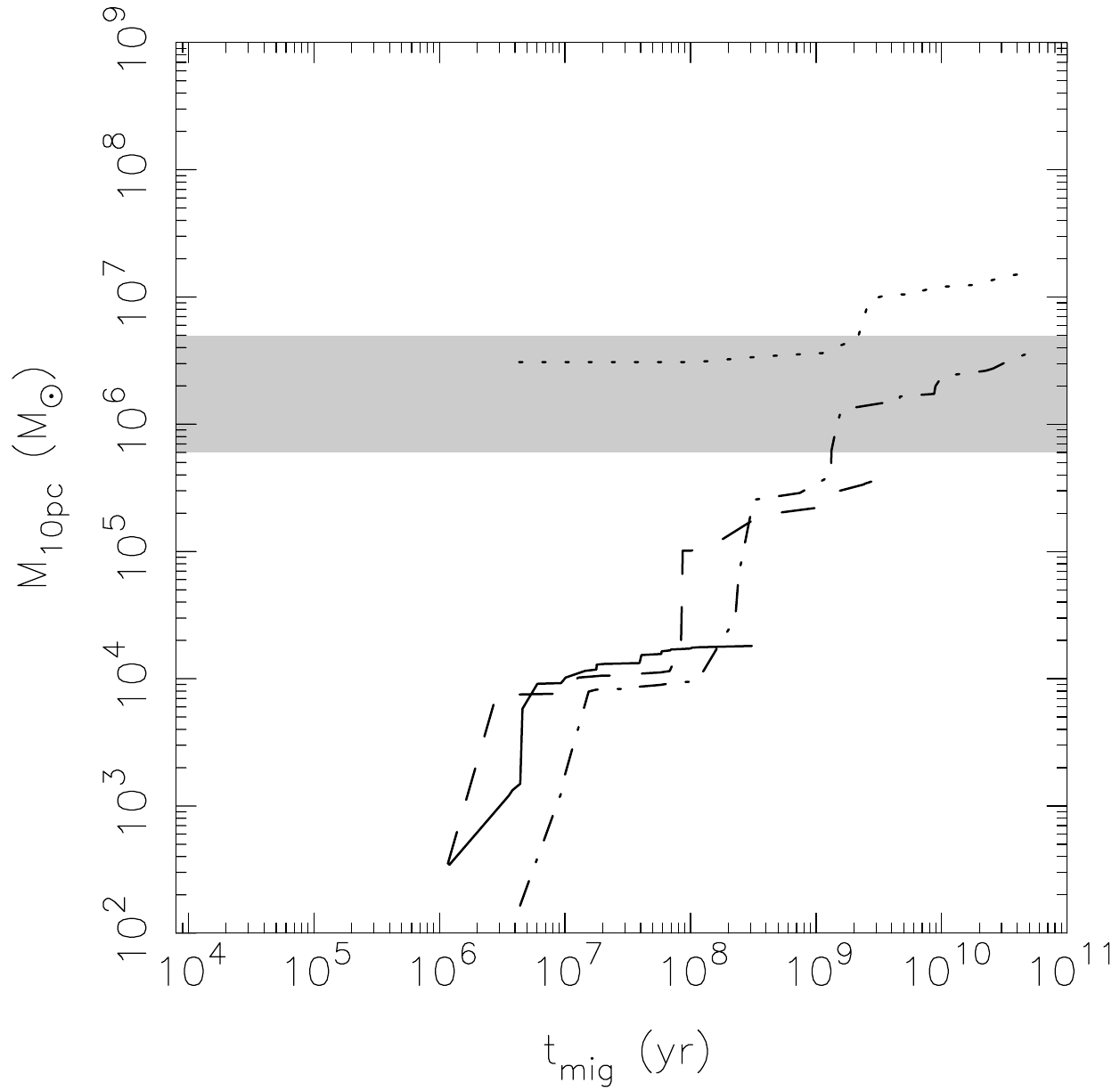}
\includegraphics[width=0.45\textwidth,clip]{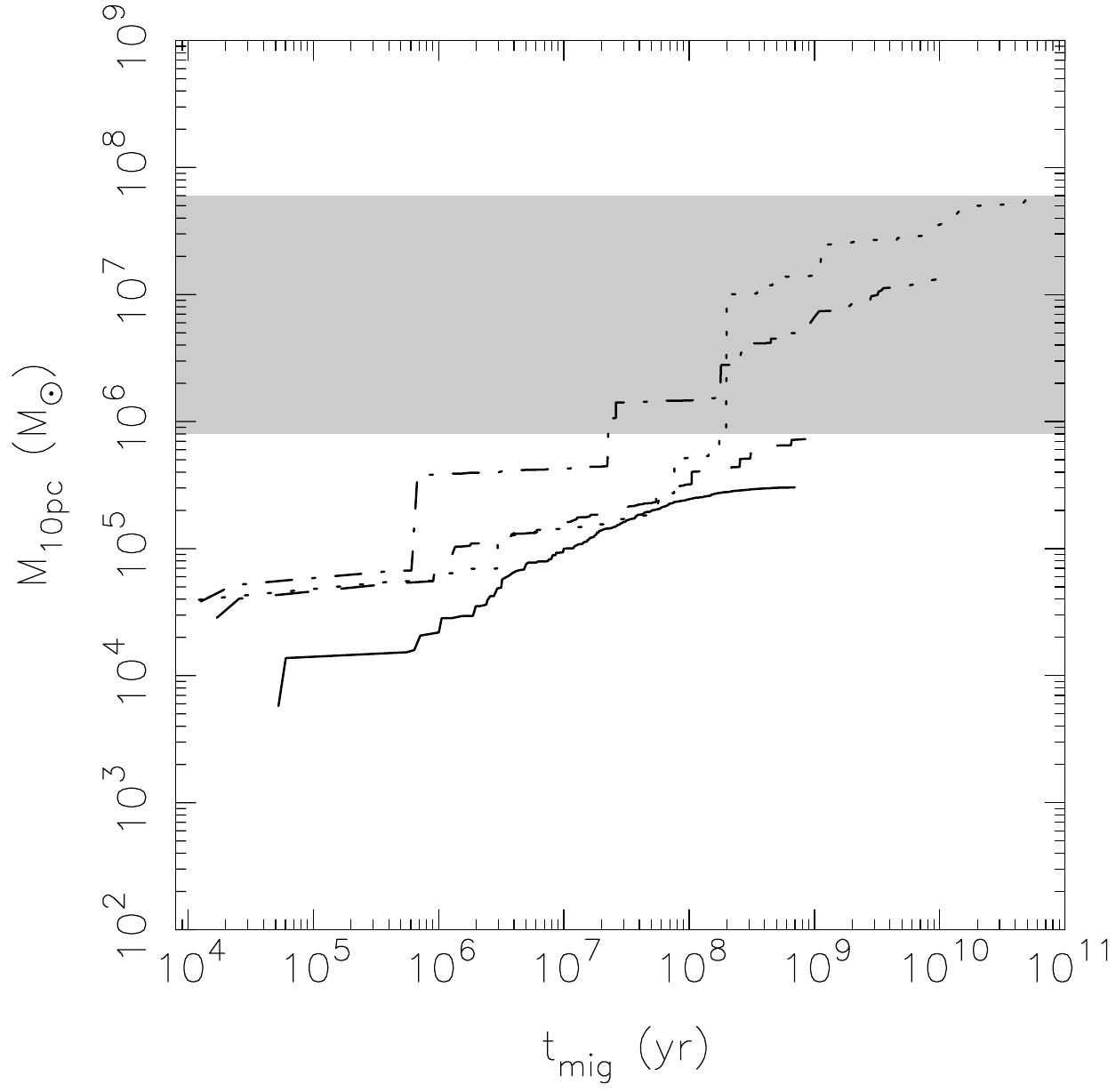}
\end{center}
\caption{The same as Figure \ref{fig:mass_time_high_z}, except that
  $90\%$ of the clusters are dissolved immediately independent of
  their mass (see Section \ref{subsec:prompt}).}
\label{fig:mass_time_high_z_prompt}
\end{figure*}

Observations of star cluster populations in nearby disk galaxies
suggest that a large ($\sim 90\%$), though still highly uncertain,
fraction of young clusters dissolve very quickly after formation,
likely as the result of an initial loosely bound state and rapid
expulsion of the natal gas cloud from the cluster by photoevaporation
and radiation pressure (this is often referred to as ``infant
mortality,'' see references cited in Section
\ref{subsec:redistribution} above).  There is evidence that the
likelihood of prompt dissolution is independent of cluster mass.  To
explore the impact of prompt dissolution on the morphological
transformation of the galaxy, we repeated the calculations carried out
in Sections \ref{subsec:sph_results} and \ref{subsec:disk_results},
but assuming that $90\%$ of randomly-selected clusters dissolve
instantaneously and deposit their entire mass at the radius at which
they formed.  If prompt dissolution is mass-dependent with, for example,
significantly less than
$90\%$ of high-mass clusters being instantaneously dissolved, then we
expect the resulting NSC mass from our simulations to be larger.  The
resulting stellar surface density profiles are shown in Figure
\ref{fig:profiles_high_z_prompt}. Now, even for our largest ICMF
truncation mass scale of $M_{\rm max}=10^7\,M_\odot$, visible
evolution of the stellar surface density profile is confined within
the inner $\sim 10-50\,\textrm{pc}$.  The time evolution of the mass
contained in the innermost $r=10\,\textrm{pc}$ is shown in Figure
\ref{fig:mass_time_high_z_prompt}.  The central masses rise as
$M_{10\,{\rm pc}}\propto t^{2/5}$ in most cases, and reach smaller
asymptotic values than without prompt dissolution. For $M_{\rm
  max}=10^4\,M_\odot$ there is no substantial departure from the
inward-extrapolated S\'ersic profile.  For $M_{\rm
  max}=(10^5-10^7)\,M_\odot$, the central masses decrease from
$\sim(10\%-20\%)$ of the values calculated without prompt dissolution
in spherical and disk galaxies.  In Tables \ref{tab:sph_prompt} and
\ref{tab:disk_prompt} we summarize the properties of the NSCs (see
Section \ref{subsec:spheroidal}) in the simulations with $90\%$ prompt
cluster dissolution and find that $M_{10\,{\rm pc}}/M_{\rm max}\sim
0.2-0.8\,\textrm{dex}$ in spheroidals and $M_{10\,{\rm pc}}/M_{\rm
  max}\sim 0.8-0.5\,\textrm{dex}$ in disks.  The timescale $t_{1/2}$
for half of the final NSC mass to accumulate extends up to
$10\,\textrm{Gyr}$.

\section{Discussion}
\label{sec:discussion}

\subsection{Comparison with Observed NSCs and Pseudobulges}
\label{subsec:observations}

Our model calculations show that cluster migration can bring about
galactic morphological transformation in which a new stellar density
component grows at the center of the galaxy.  The new component is a
drastic departure from the inward-extrapolated outer surface density
profile of the galaxy, and this calls for comparison with NSCs in
spheroidals and late-type disks, and with central light excesses in
spheroidals and pseudobulges in disks; we refer to both of the latter
phenomena as ``pseudobulges.''  The surface density profiles shown in
Figures \ref{fig:profiles_sph}, \ref{fig:profiles_disk}, and
\ref{fig:profiles_high_z_prompt} do not suggest a clear separation
into a compact component that would be compared with an NSC and a more
extended component that would be compared to a pseudobulge.  We
caution against overinterpreting the detailed profile because our
model for progressive cluster mass loss and dissolution is crude and
inevitably fails to accurately account for the evolution of the
internal structure of the cluster.  For example, if the cluster is
dense enough for its core to collapse through mass segregation and
two-body relaxation, then the cluster core could maintain integrity
longer and would migrate to smaller radii than in our calculations.
More accurate estimates of the detailed innermost stellar density
profile can only be achieved with $N$-body simulations resolving the
internal structure of the migrating clusters \citep[see,
  e.g.,][]{Capuzzo:08a,Capuzzo:08b}.  With this caveat in mind, we
pursue comparison of only the more robust characteristics of the
calculated profiles with those of observed galaxies.

\begin{deluxetable}{cccccc}
\tablecolumns{6}
\tablecaption{Spheroidal Galaxy\tablenotemark{a} NSC Properties -- $90\%$ Prompt Dissolution}
\tablehead{\colhead{Model}
& \colhead{$\log(M_{\rm max})$}
& \colhead{$\log(t_{1/2})$}
& \colhead{$\log(M_{10\,{\rm pc}})$}
& \colhead{$M_{10\,{\rm pc}}/M_{\rm sph}$}
& \colhead{$M_{10\,{\rm pc}}/M_{\rm max}$} \\
& \colhead{$(M_\odot)$}
& \colhead{(yr)}
& \colhead{$(M_\odot)$}
& \colhead{(dex)}
& \colhead{(dex)} }
\startdata
\multirow{4}{*}{High-z}
& $4$ & $6.66$ & $4.26$ & $-4.74$ & $0.26$\\
& $5$ & $8.56$ & $5.60$ & $-3.40$ & $0.60$\\
& $6$ & $9.94$ & $6.55$ & $-2.45$ & $0.55$\\
& $7$ & $9.33$ & $7.19$ & $-1.81$ & $0.19$\\
\hline
\multirow{4}{*}{Low-z}
& $4$ & $8.23$ & $4.70$ & $-4.30$ & $0.70$\\
& $5$ & $9.14$ & $5.87$ & $-3.13$ & $0.87$\\
& $6$ & $10.00$ & $6.75$ & $-2.25$ & $0.75$\\
& $7$ & $9.63$ & $7.24$ & $-1.76$ & $0.24$
\enddata
\tablenotetext{a}{$M_{\rm sph}=10^{9}\,M_\odot$, $M_{\rm
   halo}=10^{10}\,M_\odot$, no galactic mass loss.}
\label{tab:sph_prompt}
\end{deluxetable}

\begin{deluxetable}{cccccc}
\tablecolumns{6}
\tablecaption{Disk Galaxy\tablenotemark{a} NSC Properties -- $90\%$ Prompt Dissolution}
\tablehead{\colhead{Model}
& \colhead{$\log(M_{\rm max})$}
& \colhead{$\log(t_{1/2})$}
& \colhead{$\log(M_{10\,{\rm pc}})$}
& \colhead{$M_{10\,{\rm pc}}/M_{\rm disk}$}
& \colhead{$M_{10\,{\rm pc}}/M_{\rm max}$} \\
& \colhead{$(M_\odot)$}
& \colhead{(yr)}
& \colhead{$(M_\odot)$}
& \colhead{(dex)}
& \colhead{(dex)} }
\startdata
\multirow{4}{*}{High-z}
& $4$ & $7.42$ & $5.48$ & $-4.22$ & $1.48$\\
& $5$ & $8.20$ & $5.92$ & $-3.78$ & $0.92$\\
& $6$ & $8.96$ & $7.12$ & $-2.58$ & $1.12$\\
& $7$ & $9.75$ & $7.75$ & $-1.95$ & $0.75$\\
\hline
\multirow{4}{*}{Low-z}
& $4$ & $7.31$ & $5.51$ & $-4.19$ & $1.51$\\
& $5$ & $8.25$ & $6.05$ & $-3.65$ & $1.05$\\
& $6$ & $9.39$ & $7.23$ & $-2.47$ & $1.23$\\
& $7$ & $9.36$ & $7.85$ & $-1.85$ & $0.85$
\enddata
\tablenotetext{a}{$M_{\rm disk}=5\times10^{9}\,M_\odot$, $M_{\rm
   halo}=5\times10^{10}\,M_\odot$, no galactic mass loss.}
\label{tab:disk_prompt}
\end{deluxetable}

In Figures \ref{fig:mass_time_high_z} and
\ref{fig:profiles_high_z_prompt}, which show the evolution of the mass
$M_{10\,\textrm{pc}}$ contained in the innermost ten parsecs, we
indicate the range of NSC masses corresponding to the fractions of the
spheroidal luminosity contained in the NSCs in the Virgo Cluster
Survey of \citet{Ferrarese:06b} and \citet{Cote:06}, and also indicate
the range of absolute NSC masses in the survey of late-type disks by
\citet{Boker:04} and \citet{Walcher:05}.  If $M_{10\,\textrm{pc}}$ is
indeed a valid proxy for NSC mass, then this allows us to identify the
range of ICMF truncation mass scales $M_{\rm max}$ consistent with the
observed NSCs.  The models not allowing for prompt dissolution are
consistent with $M_{\rm max}\sim 10^4-10^5\,M_\odot$ for spheroidals
and $M_{\rm max}\sim 10^4-10^6\,M_\odot$ for disks.  The models with
$90\%$ prompt dissolution are consistent with $M_{\rm max}\sim
10^5-10^6\,M_\odot$ for spheroidals and $M_{\rm max} \sim
10^5-10^7\,M_\odot$ for disks.  These estimates are consistent with
the observationally inferred and theoretically anticipated values of
the ICMF truncation mass scale (Section \ref{subsec:icmf}).  The
results uniformly exclude the possibility that the ICMF truncation
mass scale in spheroidals and late-type disks is above
$10^7\,M_\odot$. 

The surface density profile modified by cluster migration already
starts departing upward from the inward-extrapolated outer S\'ersic
(or exponential) law at a radius that increases with $M_{\rm
  max}=(10^4-10^6)\,M_\odot$ from $\sim20\,\textrm{pc}$ to
$\sim300\,\textrm{pc}$ in the models without galactic mass loss and
prompt dissolution.  With galactic mass loss, the departure radius
ranges from $\lesssim10\,\textrm{pc}$ to $\sim 200\,\textrm{pc}$ for
the same range of $M_{\rm max}$.  These departure radii are relatively
small fractions of the disk exponential scale length $R_{\rm s}$ in disk
galaxies.  In the models with $90\%$ prompt dissolution, the departure
radius ranges from $\sim10\,\textrm{pc}$ to $\sim 40\,\textrm{pc}$ for
$M_{\rm max}=(10^5-10^7)\,M_\odot$.  For $M_{\rm max}=10^4\,M_\odot$,
no significant surface density excess is present.  Because of the
crudeness of the prescriptions that we employ to model cluster
migration and dissolution, and the sensitivity of the innermost
density profile to these prescriptions, we do not attempt to fit an
analytic profile, such as a two-component S\'ersic profile allowing
for a photometrically distinct central light excess (and perhaps a
third component--the NSC), to the surface density profile of the final
galaxy.  Nevertheless, the excess we observe in the central one or few
hundred parsecs is suggestive of pseudobulges that have been
identified photometrically in disk galaxies
\citep{Kormendy:04,Fisher:08,Fisher:09,Fisher:10,Weinzirl:09} and the
``central light excesses'' identified in bulgeless disks
\citep{Boker:03}, including the late-type-disk M33
\citep{Kent:87,Minniti:93}.

The effective radii of the pseudobulge components identified in the
recent surveys by Fisher et al.\ and Weinzirl et al.\ seem to be
compatible with our more optimistic models that ignore prompt
dissolution, and are on average larger than the radii within which we
detect surface density excess in the pessimistic models with $90\%$
prompt dissolution.  We caution against direct comparison because in
the present work, in an attempt to emphasize sensitivity to the
variation of the ICMF truncation mass scale $M_{\rm max}$, we have
held the parameters of our dark halo and initial baryonic disk (or
spheroid) fixed at values that seem to correspond to galaxies that are
somewhat smaller than the typical pseudobulge hosts.  We can only
conclude that a pseudobulge-like central stellar surface density
increase is generic and that cluster migration is one potential
contributor to pseudobulge assembly in disk galaxies, while other
processes, such as angular momentum transport by stellar and gaseous
bars, certainly also contribute, in line with the observation that
pseudobulge hosts generally have nuclear bars, rings, or nuclear
spirals \citep[e.g.,][]{Kormendy:04,Fisher:08}.

The most massive clusters that we have considered are still
substantially less massive than the giant $\sim10^8-10^9\,M_\odot$
clumps that are observed to be present and are theoretically expected
to be forming in globally gravitationally unstable,
rapidly-star-forming massive disks at high redshift \citep[e.g.,][and
  references
  therein]{Noguchi:99,Bournaud:07,Elmegreen:08b,Dekel:09b,Tacconi:10}. The
super star clusters forming in these giant clumps should be more
immune to dissolution in the tidal field of the galaxy and could reach
the galactic central region intact. We speculate that there could be a
critical characteristic ICMF mass scale above which clusters migrate
intact and merge to produce a classical bulge \citep[see,
  e.g.,][]{Immeli:04,Elmegreen:08b,Ceverino:10}, and below which they
suffer substantial mass loss en route to the galactic center and thus
give rise to a pseudobulge.

\subsection{Implications for Massive Black Holes}
\label{subsec:mbh}

The apparent agreement of NSC-mass-to-galactic-stellar-mass ratios in
spheroidals ($\sim2\times10^{-3}$;
\citealt{Cote:06,Ferrarese:06b,Wehner:06}) and
massive-black-hole-to-galactic stellar mass ratios in ellipticals and
bulges ($\sim[1-2]\times10^{-3}$; e.g.,
\citealt{Kormendy:95,Wandel:99,Kormendy:01,Merritt:01a,McLure:02,Marconi:03,Haring:04})
has prompted speculation that the same process may be responsible for
the formation of NSCs and black holes.  While the formation and growth
of a massive black hole undoubtedly requires a gas-dynamical,
dissipative process, our results suggest that an NSC can be assembled
nondissipatively, by the collisionless migration of star clusters, and
thus, the observed agreement could be a coincidence.

Although the well-studied NSC host galaxy M33 does not contain a
central massive black hole \citep[][and references
  therein]{Merritt:01b,Gebhardt:01}, another one, NGC\,4395, does
\citep{Filippenko:03}, and still others contain AGNs \citep{Seth:08}.
Since AGNs and the growth of a central black hole require gas inflow
into the center of the galaxy, NSC growth from migrating disk clusters
would not generally be accompanied with black hole growth, because
disk clusters contribute stellar mass without augmenting the black
hole mass (although gas inflow may be enhanced by the migrating
clusters, see, e.g., \citealt{Goodman:01,Chang:08}). This suggests
that the central black hole mass in bulgeless disks should not be
correlated with the mass of the NSC and that any pseudobulge that is
present, unless the migrating clusters independently synthesize
massive (or intermediate-mass) black holes which, in this case, they
would deliver to the center to merge to form a more massive central
black hole \citep{Elmegreen:08a}.

\pagebreak

\section{Conclusions}
\label{sec:conclusions}

We studied the morphological transformation of isolated spheroidal and
late-type disk galaxies driven by migration and tidal disruption of
embedded star clusters.  With the help of empirically and
theoretically calibrated cluster migration and disruption time scales,
we have tracked the mass that accumulates in the central region of a
spheroidal or late-type disk galaxy and have compared the resulting
innermost surface density profile to that of NSC and pseudobulge host
galaxies.  We have focused on the variation in the degree of
galaxy transformation with the ICMF truncation mass scale; the
assumptions and limitations of our model are described in Section
\ref{subsec:limitations}.  Our main conclusions are:

1. The amount stellar mass that the migrating clusters transport into
the central few tens or hundreds of parsecs is sensitive to the
maximum mass of the clusters forming in the galaxy.  Larger ICMF
high-mass truncation mass, $M_{\rm max}$, scales yield larger and more
spatially extended central stellar accumulations.

2. Because our model for progressive cluster disruption is crude, we
cannot attach significance to the detailed shape of the innermost
surface density profile (the number of photometrically distinct
components, inflection points, etc.), but we do compare the robust
features of our synthetic profiles with the observed properties of NSC
host galaxies.

3. Our model yields NSC masses compatible with the observed NSC masses
and scaling relations if the ICMF truncation mass scale is
$10^5\,M_\odot\lesssim M_{\rm max}\lesssim 10^6\,M_\odot$, which is
consistent with direct $M_{\rm max}$ estimates in the nearby
star-forming galaxies.  To match NSC masses in galaxies with a higher
degree of prompt dissolution (``infant mortality''), higher ICMF
truncation mass scales are needed.

4. The surface density profiles modified by cluster migration exhibit
excesses above the inward-extrapolated outer S\'ersic (or exponential)
profile in the central few tens to few hundreds of parsecs.  These
excesses are suggestive of the pseudobulge phenomenon in disk
galaxies. \emph{The formation of NSCs and central surface density
  excesses in bulgeless galaxies is inevitable and generic simply as a
  consequence of the clustered nature of star formation.}

\acknowledgements

We thank D. Fisher, M. Gieles, J. Kormendy, and C. J. Walcher for
invaluable discussions and also thank the organizers of the workshop
{\it Nuclear Star Clusters across the Hubble Sequence} at MPIA,
Heidelberg, 2008.  We would also like to thank an anonymous referee
for helpful comments on an earlier version of this work.  We
acknowledge NSF grant AST-0708795.

\end{document}